\documentclass{article}

 \usepackage[eandd, final]{neurips_2026}

\usepackage[utf8]{inputenc} 
\usepackage[T1]{fontenc}    
\usepackage{hyperref}       
\usepackage{url}            
\usepackage{booktabs}       
\usepackage{amsfonts}       
\usepackage{nicefrac}       
\usepackage{microtype}      
\usepackage{xcolor}         
\usepackage{graphicx}
\usepackage{amsmath}
\usepackage{pifont}
\usepackage{multirow}
\usepackage{array}
\usepackage{wrapfig}
\usepackage{caption}
\usepackage[table]{xcolor}
\usepackage{makecell}
\usepackage{pifont}
\usepackage{adjustbox}
\usepackage{listings}
\usepackage{tabularx}
\usepackage{longtable}
\usepackage{placeins}
\usepackage{afterpage}
\definecolor{groupblue}{RGB}{235,242,252}
\newcolumntype{C}[1]{>{\centering\arraybackslash}m{#1}}
\newcommand{\cmark}{\textcolor{green!60!black}{\ding{51}}}
\newcommand{\xmark}{\textcolor{red}{\ding{55}}}
\newcolumntype{Y}{>{\raggedright\arraybackslash}X}
\newcommand{\rowrule}{%
  \arrayrulecolor{black!18}\specialrule{0.25pt}{0pt}{0pt}\arrayrulecolor{black}%
}
\lstdefinestyle{promptstyle}{
  basicstyle=\footnotesize\ttfamily,
  breaklines=true,
  breakatwhitespace=false,
  columns=fullflexible,
  keepspaces=true,
  showstringspaces=false,
  linewidth=\linewidth,
  postbreak=\mbox{\textcolor{gray}{$\hookrightarrow$}\space}
}
\title{ProSearch: Benchmarking Multi-Constraint Protocol Retrieval in Experimental Science }

\author{
\textbf{Wenliang Liang\textsuperscript{1,3}\thanks{Equal contribution.}
\quad
Zhiyuan Ning\textsuperscript{4,*}\thanks{Project lead.}
\quad
Haolin Chen\textsuperscript{2,3}}
\\[0.5em]
\textbf{Yuanchun Zhou\textsuperscript{1,3}
\quad
Wenjuan Cui\textsuperscript{1,3}
\quad
Yi Du\textsuperscript{1,3}\thanks{Corresponding author.}}
\\[0.6em]
\textsuperscript{1} Computer Network Information Center, Chinese Academy of Sciences
\\
\textsuperscript{2} Institute of Zoology, Chinese Academy of Sciences
\\
\textsuperscript{3} University of Chinese Academy of Sciences
\\
\textsuperscript{4} Westlake University
\\[0.5em]
\texttt{ningzhiyuan@westlake.edu.cn}
\\
\texttt{chenhaolin23@mails.ucas.ac.cn}
\\
\texttt{\char123 wlliang,zyc,wenjuancui,duyi\char125 @cnic.cn}
}

\begin{document}

\maketitle

\begin{abstract}
  In AI for Science scenarios, experimental design and method reproduction require retrieving protocols that satisfy multiple interdependent constraints, yet existing benchmarks lack systematic evaluation of this capability. To address this gap, we introduce ProSearch, a benchmark for multi-constraint protocol retrieval in experimental science. ProSearch has three key features:  (1) realism, with queries grounded in real experimental records and expert-defined templates; (2) complexity, with 60 experimental fields and their dependencies; and (3) diagnosticity, with 3–5 explicit constraints per query, including numerical and negation constraints. We construct ProSearch through LLM-based extraction, data-driven constraint selection, controlled query generation, and evidence-based relevance validation. The resulting benchmark contains 1,515 queries across six experimental science domains.
  We conduct extensive experiments on ProSearch with 19 representative retrievers and complementary analyses. The best-performing retriever achieves an average nDCG@10 of only 0.290, which improves to 0.474 with reranking. Structured diagnostic analysis shows that explicit constraint matching provides a strong ranking signal, while manual error analysis identifies joint constraint binding as a key bottleneck. ProSearch provides a testbed for systematically evaluating scientific retrieval systems. \footnote{
We provide our code in \href{https://github.com/kg4sci/ProSearch}{https://github.com/kg4sci/ProSearch}, data in \href{https://huggingface.co/datasets/kg4sci/ProSearch}{https://huggingface.co/datasets/kg4sci/ProSearch}.}
\end{abstract}

\section{Introduction}

As AI for Science advances, artificial intelligence systems are increasingly being used for scientific tasks~\citep{xie2025far,weng2026deepreviewer} such as experimental design~\citep{ma2025preference,wang2026microbediscover} and protocol reproduction~\citep{gridach2025agentic,chen2025ai4research}. In these settings, systems must not only understand the research question but also retrieve actionable methodological evidence from the scientific literature to identify protocols that meet specific experimental requirements~\citep{kasmaee2025chembed}. In this context, protocol retrieval has become an important foundational capability for AI for Science~\citep{kon2025exp,eger2025transforming,gupta2022matscibert,gu2021domain}.

Beyond topical relevance, protocol retrieval aims to identify method paragraphs describing experimental protocols that satisfy multiple specific conditions~\citep{liu2025bioprobench}. For example, a researcher may require a specific cell line to be cultured at $37^\circ$C while excluding the use of DMEM as the culture medium. Retrieval systems must not only identify these conditions but also verify that they hold within the same coherent experimental configuration. However, method paragraphs often describe multiple experimental configurations or variants within the same context~\citep{sun2025unleashing}. Therefore, even when all query conditions appear in the same paragraph, they may correspond to different experimental procedures, leading keyword-based retrieval methods to produce false matches.

Existing benchmarks provide limited evaluation of this capability~\citep{Zhong2025BenchmarkingRG,ajith2024litsearch,chen2026prereviewbench}. Traditional scientific datasets focus on abstract-level fact verification or passage-level question answering~\citep{wadden2020fact,Jin2019PubMedQAAD}. More recent reasoning-intensive benchmarks, such as BRIGHT~\citep{Su2024BRIGHTAR}, introduce queries that require inference beyond surface-level matching, but do not impose explicit multi-constraint requirements. MultiConIR~\citep{Lu2025MultiConIRTM} addresses multi-constraint retrieval, yet focuses on structured items rather than unstructured scientific text. Existing evaluations do not adequately measure whether retrievers can jointly bind multiple constraints to a single experimental variant.

However, constructing such a benchmark presents several challenges~\cite{song2026expert}. First, realistic protocol queries are not readily available in existing retrieval datasets and cannot be generated by simply combining keywords. They must reflect meaningful experimental needs and follow valid experimental logic. Moreover, a plausible query does not necessarily have supporting evidence in the retrieval corpus. Query construction must therefore be grounded in real experimental protocols. Finally, relevance labeling requires verifying that the retrieved text supports all query constraints within a consistent experimental configuration, making reliable annotation difficult to automate.

To address these challenges, we develop an expert-guided construction pipeline. Domain experts define experimental fields, dependencies, and query templates. We instantiate these templates using values from extracted protocol records and select constraint combinations in the corpus. LLM-assisted filtering and query generation then produce meaningful natural-language queries. For relevance labeling, we use structured records to identify candidate paragraphs but determine final relevance from the original text, labeling a paragraph as relevant only when a single experimental configuration supports all query constraints. Through this pipeline, we introduce \textbf{ProSearch}, a multi-constraint protocol retrieval benchmark comprising 1,515 queries across six experimental science domains. ProSearch has three key characteristics: (1) \textit{realism}, with queries grounded in real experimental records and expert-defined templates; (2) \textit{complexity}, with 60 experimental fields and their dependencies; and (3) \textit{diagnosticity}, with 3–5 explicit constraints per query, including numerical and negation constraints.

To assess current retrieval capabilities on ProSearch, we conduct a series of comprehensive experiments. Our evaluation of 19 retrievers shows that ProSearch is a major challenge; even the best models achieve an average nDCG@10 of only 0.290, leaving large room for improvement. We find that reranking improves this score to 0.474, while query reformulation yields only modest gains. Structured diagnostic analysis based on extracted protocol records achieves an average nDCG@10 of 0.774, increasing to 0.814 when dense similarity is used to break ties. Manual error analysis further reveals that 90\% of sampled top-ranked non-relevant paragraphs violate at least one query constraint, highlighting constraint binding as a key bottleneck.  Overall, we hope ProSearch serves as a testbed for building both better retrievers and next generation scientific agents that can truly handle complex experimental logic.

\section{Related Work}

\paragraph{Scientific retrieval benchmarks.} As retrieval systems increasingly support scientific research~\citep{pan2026beyond}, benchmarks have been developed to evaluate a broad range of tasks.
SCIDOCS~\citep{Cohan2020SPECTERDR} and SciRepEval~\citep{singh2023scirepeval} evaluate scientific document representations.
SciFact~\citep{wadden2020fact} focuses on scientific claim verification, while PubMedQA~\citep{Jin2019PubMedQAAD} and BioASQ~\citep{nentidis2025overview} address biomedical question answering.
TREC-COVID~\citep{Roberts2020TRECCOVIDRA} and LitSearch~\citep{ajith2024litsearch} evaluate scientific literature retrieval.
Other work focuses more specifically on the relevance of scientific methods.
CSFCube~\citep{BENCHMARKS2021_20f07591} evaluates scientific paper relevance based on research objectives, methods, and results.
MIR~\citep{garikaparthi2025mir} focuses on retrieving papers that can provide methodological inspiration for a research problem.
However, methodological relevance does not necessarily mean that specific experimental conditions are satisfied.
ProSearch focuses on method paragraphs and evaluates whether they provide evidence that jointly satisfies multiple experimental requirements.

\begin{table}[t]
\centering
\small
\setlength{\tabcolsep}{5pt}
\renewcommand{\arraystretch}{1.08}
\caption{
Comparison of ProSearch with representative retrieval benchmarks.
\textit{Sci.} denotes scientific IR, and \textit{Evidence Unit} denotes retrieval granularity.
For \textit{Query Construction}, \textit{Hum.}, \textit{LLM}, and \textit{Constr.} denote human involvement, LLM use, and explicit multi-constraint queries.
For \textit{Relevance Requirement}, \textit{Hard}, \textit{Bind.}, and \textit{Proc.} denote mandatory satisfaction of all query requirements, joint grounding of requirements to the same target context, and process, mechanism, or workflow-level reasoning, respectively.
}
\label{tab:benchmark_comparison}
\resizebox{\textwidth}{!}{
\begin{tabular}{l r c l c c c c c c}
\toprule
\multirow{2}{*}{Benchmark}
& \multirow{2}{*}{\# Queries}
& \multirow{2}{*}{Sci.}
& \multirow{2}{*}{Evidence Unit}
& \multicolumn{3}{c}{Query Construction}
& \multicolumn{3}{c}{Relevance Requirement} \\
\cmidrule(lr){5-7}
\cmidrule(lr){8-10}
&
&
&
& Hum.
& LLM
& Constr.
& Hard
& Bind.
& Proc. \\
\midrule
MultiConIR~\citep{Lu2025MultiConIRTM}
& 23,070
& \xmark
& Structured Item
& \xmark & \cmark & \cmark
& \cmark & \cmark & \xmark \\

BRIGHT~\citep{Su2024BRIGHTAR}
& 1,384
& \xmark
& Document
& \cmark & \xmark & \xmark
& \xmark & \xmark & \cmark \\

FollowIR~\citep{weller2025followir}
& 102
& \xmark
& Document
& \cmark & \xmark & \xmark
& \cmark & \xmark & \xmark \\

\midrule
TREC-COVID~\citep{Roberts2020TRECCOVIDRA}
& 50
& \cmark
& Document
& \cmark & \xmark & \xmark
& \xmark & \xmark & \xmark \\

SciFact~\citep{wadden2020fact}
& 1,409
& \cmark
& Abstract
& \cmark & \xmark & \xmark
& \xmark & \xmark & \xmark \\

BioASQ-13b~\citep{nentidis2025overview}
& 5,729
& \cmark
& Passage
& \cmark & \xmark & \xmark
& \xmark & \xmark & \cmark \\

ChemRAG-Bench~\citep{Zhong2025BenchmarkingRG}
& 1,932
& \cmark
& Passage
& \cmark & \xmark & \xmark
& \xmark & \xmark & \cmark \\

LitSearch~\citep{ajith2024litsearch}
& 597
& \cmark
& Document
& \cmark & \cmark & \xmark
& \xmark & \xmark & \xmark \\

\midrule
ProSearch
& 1,515
& \cmark
& Method Paragraph
& \cmark & \cmark & \cmark
& \cmark & \cmark & \cmark \\
\bottomrule
\end{tabular}
}
\end{table}

\paragraph{Constraint-aware and reasoning-intensive retrieval.}
Recent benchmarks evaluate whether systems can handle requirements beyond topical relevance.
FollowIR~\citep{weller2025followir} and related instruction-following benchmarks~\citep{Song2025IFIRAC,Zhou2024BeyondCR} study whether retrievers can follow query-level constraints.
MultiConIR~\citep{Lu2025MultiConIRTM} studies multi-constraint retrieval over structured items, while Box Embeddings~\citep{dasgupta2023answering} explore attribute intersections and negation.
BRIGHT~\citep{Su2024BRIGHTAR} focuses on reasoning-intensive retrieval where relevant documents require inference beyond surface matching.
ProSearch differs by emphasizing joint constraint binding within unstructured experimental methods text.
A method paragraph may describe multiple experimental configurations.
Even when all parameters required by a query appear in the same paragraph, they may belong to different experimental configurations.
ProSearch therefore requires all query constraints to be jointly satisfied within the same experimental configuration.
Table~\ref{tab:benchmark_comparison} compares representative retrieval benchmarks in terms of retrieval units, query construction, and relevance requirements.

\section{ProSearch}

In this section, we present ProSearch, a multi-constraint protocol retrieval benchmark for experimental science. We first define the protocol retrieval task (Section ~\ref{section3.1}). Then, we detail our expert-guided construction pipeline (Section ~\ref{section3.2}). Finally, we evaluate the benchmark's quality through expert assessments (Section ~\ref{section3.3}). Table~\ref{tab:data_statistics} summarizes the benchmark statistics.

\subsection{Task Formulation}
\label{section3.1}

Given a query $q$ and a retrieval corpus $\mathcal{D}=\{p_1,\ldots,p_n\}$, where each $p_i$ is a method paragraph, the goal is to rank paragraphs according to whether they satisfy the experimental requirements specified by $q$.
Each query is associated with a set of protocol constraints $\mathcal{C}_q=\{c_1,\ldots,c_m\}$.
Each constraint is represented as $c_j=(f_j,o_j,v_j)$, where $f_j$ denotes a domain-specific protocol field, $o_j$ denotes the operator, and $v_j$ denotes the required value.
Table~\ref{tab:protocol_schema_examples} illustrates how protocol fields, operators, and values are instantiated as query constraints.

A method paragraph may describe multiple experimental configurations.
Therefore, we define paragraph relevance at the protocol-record level.
Let $\mathcal{R}(p)=\{r_1,\ldots,r_k\}$ denote the set of protocol records described in paragraph $p$.
Each record corresponds to one coherent experimental configuration and does not combine evidence across different configurations in the same paragraph.
A paragraph $p$ is relevant to query $q$ if and only if there exists at least one protocol record $r\in\mathcal{R}(p)$ that satisfies all constraints in $\mathcal{C}_q$:
\begin{equation}
p \in \mathcal{D}^{+}_{q}
\Longleftrightarrow
\exists r \in \mathcal{R}(p), \ \forall c_j \in \mathcal{C}_q,\ r \models c_j.
\end{equation}

\begin{table*}[!t]
\centering
\small
\setlength{\tabcolsep}{4pt}
\renewcommand{\arraystretch}{1.08}
\caption{
Data statistics of ProSearch.
$\mathcal{Q}$ denotes the number of queries, $\mathcal{D}$ denotes the number of paragraphs, and $\mathcal{D}^{+}$ denotes the average number of positive paragraphs per query.
Average lengths are computed using GPT-2 tokens.
\#Field Templates denotes the number of field combinations used to construct queries.
Top-10 Coverage denotes the fraction of queries covered by the ten most frequent field templates.
}
\label{tab:data_statistics}
\resizebox{\textwidth}{!}{
\begin{tabular*}{\textwidth}{@{\extracolsep{\fill}}lcccccccc@{}}
\toprule
& \multicolumn{3}{c}{Total Number}
& \multicolumn{2}{c}{Avg. Length}
& \multicolumn{2}{c}{Query Diversity}
& \multirow{2}{*}{Examples} \\
\cmidrule(lr){2-4}
\cmidrule(lr){5-6}
\cmidrule(lr){7-8}
Domain
& $\mathcal{Q}$
& $\mathcal{D}$
& $\mathcal{D}^{+}$
& $\mathcal{Q}$
& $\mathcal{D}$
& \#Field Templates
& Top-10 Coverage
& \\
\midrule
Cell Culture         & 284 & 46,004 & 7.1 & 22.0 & 218.7 & 14  & 97.2\% & Tab.~\ref{tab:example_cell_culture} \\
Immunotechnology     & 264 & 93,315 & 9.0 & 26.9 & 228.1 & 26  & 85.6\% & Tab.~\ref{tab:example_immunotechnology} \\
Electrochemistry     & 201 & 91,375 & 8.7 & 28.0 & 231.4 & 44  & 58.7\% & Tab.~\ref{tab:example_electrochemistry} \\
Organic Synthesis    & 252 & 30,371 & 9.8 & 26.1 & 262.2 & 57  & 49.2\% & Tab.~\ref{tab:example_organic_synthesis} \\
Inorganic Materials  & 247 & 46,136 & 4.0 & 27.2 & 217.7 & 104 & 34.8\% & Tab.~\ref{tab:example_inorganic_materials} \\
Polymer Materials    & 267 & 50,977 & 4.7 & 25.4 & 228.4 & 37  & 70.4\% & Tab.~\ref{tab:example_polymer_materials} \\
\bottomrule
\end{tabular*}}
\end{table*}

\subsection{Dataset Construction}
\label{section3.2}

\begin{figure*}[t]
    \centering
    \includegraphics[width=\textwidth]{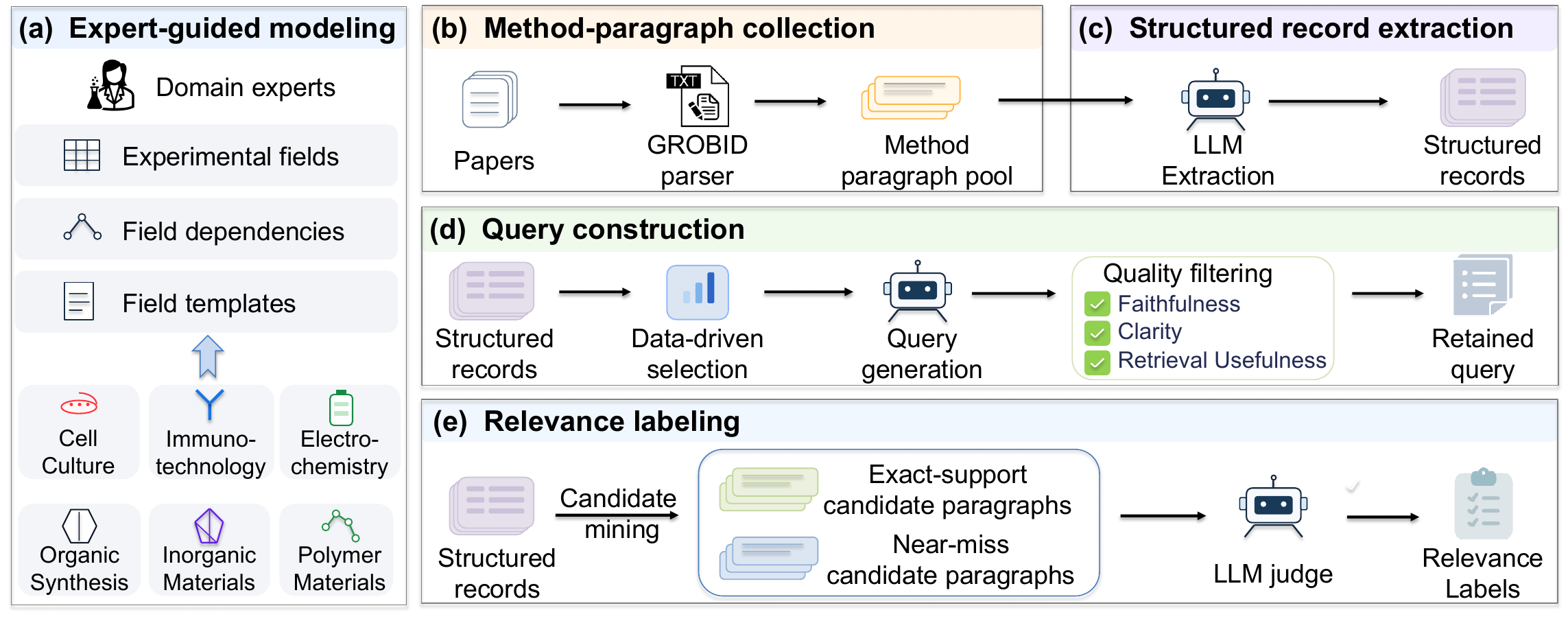}
    \caption{
    ProSearch data construction pipeline.
    }
    \label{fig:data_construction}
\end{figure*}

In this section, we present the construction of ProSearch. Our pipeline (Figure~\ref{fig:data_construction}) is designed to ensure: (1) \textbf{Retrieval Usefulness}, where queries represent meaningful requests that reflect the actual information needs of domain researchers; and (2) \textbf{Verifiability}, where paragraph relevance is determined by explicit, evidence-grounded validation.

\paragraph{Expert-Guided Modeling.} To construct meaningful multi-constraint protocol queries, we first collaborate with domain experts to define experimental fields and their usage rules, then organize these fields through shared roles and query templates.

\textit{Domain Coverage and Field Selection.}
We cover six domains: Cell Culture, Immunotechnology, Electrochemistry, Organic Synthesis, Inorganic Materials, and Polymer Materials.
For each domain, an expert identifies a representative experiment type and selects fields according to three criteria:
(1) \textit{Experimental relevance} requires a field to describe a key aspect of the target protocol, such as cell type or medium in cell culture.
(2) \textit{Textual extractability and standardizability} requires that field values can be identified in method paragraphs and normalized under controlled rules.
(3) \textit{Discriminative power} refers to the ability to distinguish protocols with similar goals but different configurations. For example, temperature alone provides little discrimination when most protocols use 37~$^\circ$C.
Experts also specify each field's value type, permitted operators, dependencies, and usage rules.

\textit{Role-based Template Design.}
To organize fields consistently across domains, we classify them into four functional roles.
\textit{Goal} denotes the primary experimental target anchoring each query template, such as the target cell type in cell culture.
\textit{Input} covers materials, samples, or reagents; \textit{Condition} covers experimental parameters such as temperature, concentration, time, or atmosphere; and \textit{Method} covers procedural operations, treatments, or measurements.
We define 20 shared role-level templates (Appendix Table~\ref{tab:role_level_templates}), each containing 3--5 role slots.
These role-level templates are instantiated with domain-specific fields to generate candidate field-level templates.

\textit{Field-level Template Filtering.}
Not every field combination generated from a role-level template forms a meaningful retrieval request.
We therefore filter field-level templates using the expert-defined dependencies and usage rules.
For example, the role-level template \textit{Goal + Input + Condition} can produce \textit{Cell Type + Basal Medium + Confluence}. This combination is filtered out because Confluence depends on Culture Mode. In contrast, \textit{Cell Type + Basal Medium + Confluence + Culture Mode}, generated from \textit{Goal + Input + Condition + Method}, satisfies the dependency and is retained.
Complete field definitions, rules, and templates are provided in Appendix~\ref{app:expert_guided_protocol_modeling}.

\paragraph{Corpus Collection and Paragraph Filtering.} 
For each domain, we collect English open-access research articles published between 2020 and 2025 via OpenAlex~\citep{Priem2022OpenAlexAF}.
We use GROBID~\citep{Lopez2009GROBIDCA} to reconstruct section structures and paragraph boundaries.
We identify method-related sections by matching top-level section headings against expert-defined patterns and retain paragraphs from these sections.
We exclude non-paragraph content, such as tables and figure captions, and discard paragraphs exceeding 2,048 GPT-2 tokens.
The retained method paragraphs form the retrieval corpus $\mathcal{D}$.
Further details are provided in Appendix~\ref{app:corpus_filtering}.

\paragraph{Protocol Record Construction.}
A method paragraph may describe multiple experimental configurations, such as different material combinations or reaction conditions.
To avoid mixing conditions from different configurations, we use GPT-5.2 to extract a separate structured record for each configuration based on expert-defined experimental fields. 
Each record preserves the field values and their supporting text spans from the original paragraph, grounding the extraction in textual evidence. Since the same experimental value may be expressed differently across papers, we then use GPT-5.2 to normalize the extracted field values~\citep{ai2025distilling}. 
For closed-set fields, we select the most specific predefined category supported by the text.
For open-set fields, we normalize synonymous expressions while preserving the specific identities supported by the original text.
Further details are provided in Appendix~\ref{app:prompts_examples}.

\paragraph{Query Construction.}
We instantiate the filtered field-level templates with values from a single normalized protocol record to generate candidate constraint sets. For each set, we compute its support count using the extracted records, counting a paragraph once if at least one of its records satisfies all constraints. We retain sets with support counts between 3 and 8 to balance task specificity and evaluation stability, excluding overly broad constraint sets and those with insufficient structured support. We then sample constraint sets stratified by constraint count and use GPT-5.4 to filter out scientifically uninformative sets and convert the retained ones into natural-language queries~\citep{cai2023resolving}. We use Claude Sonnet 4.6 to evaluate the queries along three dimensions:
(1) \textbf{Faithfulness}, ensuring no constraints are added or omitted;
(2) \textbf{Clarity}, ensuring natural and unambiguous expression; and (3) \textbf{Retrieval Usefulness}, confirming that the query represents a meaningful scientific request.
Only queries passing all three checks are retained. This process yields 1,551 constructed queries across the six domains.
Figure~\ref{fig:query_composition} summarizes the resulting query patterns.
Detailed sampling statistics and filtering examples are provided in
Appendix~\ref{app:query_generation_details}.

\paragraph{Relevance Labeling.} Since exhaustive labeling is impractical, we use structured records to construct an annotation pool of potentially relevant paragraphs for each query. We first identify paragraphs containing at least one extracted record matching all query constraints as \textit{exact-support candidates}.
Extraction omissions or normalization inconsistencies may cause exact matching to miss relevant paragraphs.
We therefore also include \textit{near-miss candidates}: paragraphs containing at least one extracted record matching all but one constraint.
We sample up to 100 near-miss candidates per query.
Both candidate types are defined by structured matching over extracted records and do not determine relevance labels.
We then use GPT-5.4 to judge candidate relevance based on the original paragraph text. 
The model examines experimental configurations separately and labels a paragraph as relevant only when the text supports all query constraints within the same configuration.
Of the 1,551 constructed queries, 36 had no associated relevance judgments (qrels) and were excluded from retrieval evaluation, leaving 1,515 queries in the final evaluation set.
More details are provided in Appendix~\ref{app:prompts_examples}.

\subsection{Data Quality Evaluation}
\label{section3.3}

\paragraph{Query Quality.}

\begin{wraptable}{r}{0.38\textwidth}
\vspace{-2mm}
\centering
\small
\setlength{\tabcolsep}{6pt}
\renewcommand{\arraystretch}{1.08}
\captionof{table}{Expert evaluation of query quality on 120 sampled queries.}
\label{tab:query_assessment}
\begin{tabular*}{\linewidth}{@{\extracolsep{\fill}}lc@{}}
\toprule
Criterion & Pass Rate \\
\midrule
Faithfulness         & 100\% \\
Clarity              & 100\% \\
Retrieval Usefulness & 91.7\% \\
\bottomrule
\end{tabular*}
\vspace{-2mm}
\end{wraptable}

We draw a stratified sample of 120 queries for expert evaluation, including 45 from Cell Culture and 15 from each of the other five domains. As shown in Table~\ref{tab:query_assessment}, all sampled queries are faithful to their structured constraints and clearly expressed, while 91.7\% are judged useful for retrieval. The remaining cases mainly involve insufficient experimental context, such as electrochemical queries asking for applied potential without specifying the material or reaction being studied.

\begin{table*}[ht]
\centering
\small
\renewcommand{\arraystretch}{1.08}

\begin{minipage}[t]{0.32\textwidth}
\vspace{0pt}
\begin{minipage}[t][2.0cm][t]{\linewidth}
\captionof{table}{
Expert agreement with relevance labels across six domains.
The audit includes 60 pairs from Cell Culture and
45 from each other domain.
}
\label{tab:label_quality}
\end{minipage}

\centering
\setlength{\tabcolsep}{5pt}
\begin{tabular*}{\linewidth}{@{\extracolsep{\fill}}lc@{}}
\toprule
Domain & Accuracy \\
\midrule
Cell Culture        & 91.7\% \\
Electrochemistry    & 95.6\% \\
Immunotechnology    & 88.9\% \\
Inorganic Materials & 80.0\% \\
Organic Synthesis   & 93.3\% \\
Polymer Materials   & 84.4\% \\
\midrule
Overall             & 89.1\% \\
\bottomrule
\end{tabular*}
\end{minipage}
\hfill
\begin{minipage}[t]{0.64\textwidth}
\vspace{0pt}
\begin{minipage}[t][2.0cm][t]{\linewidth}
\captionof{table}{
Annotation pool statistics in ProSearch.
Exact-Support and Near-Miss report the percentages of
candidates labeled relevant in each group.
NM Contrib. denotes the fraction of all identified
positives originating from near-miss candidates.
}
\label{tab:candidate_pool_quality}
\end{minipage}

\centering
\setlength{\tabcolsep}{4pt}
\begin{tabular*}{\linewidth}{@{\extracolsep{\fill}}lccc@{}}
\toprule
Domain & Exact-Support & Near-Miss & NM Contrib. \\
\midrule
Cell Culture        & 95.6\% & 4.3\%  & 52.9\% \\
Electrochemistry    & 83.1\% & 9.5\%  & 70.0\% \\
Immunotechnology    & 91.3\% & 6.8\%  & 69.7\% \\
Inorganic Materials & 90.5\% & 4.7\%  & 30.4\% \\
Organic Synthesis   & 87.1\% & 17.2\% & 72.0\% \\
Polymer Materials   & 90.0\% & 5.6\%  & 37.6\% \\
\midrule
Overall              & 90.1\% & 7.5\%  & 59.8\% \\
\bottomrule
\end{tabular*}
\end{minipage}

\end{table*}

\paragraph{Label Quality}
We independently audit 285 query--paragraph pairs across six domains. As shown in Table~\ref{tab:label_quality}, expert judgments agree with the GPT-5.4 relevance labels in 89.1\% of cases overall. For Immunotechnology, independent double annotation yielded Cohen's $\kappa=0.776$. To assess validator sensitivity, we sample 300 queries (50 per domain) and re-judge all 16,700 pairs in their original annotation pools using Qwen3-235B and Claude Sonnet 4.6. Their agreements with GPT-5.4 are 66.5\% and 90.2\%, respectively. Despite this variation in label agreement, the resulting retriever rankings remain highly correlated with those obtained using GPT-5.4 labels, with Spearman's $\rho=0.970$ and $0.982$, respectively. 

\paragraph{Annotation Pool Analysis.}
Table~\ref{tab:candidate_pool_quality} reports the relevance judgments for the two candidate groups. Exact-Support candidates have a positive rate of 90.1\%, while 7.5\% of Near-Miss candidates are also judged relevant. Notably, Near-Miss candidates account for 59.8\% of all identified positives, demonstrating their importance in recovering relevant paragraphs missed by exact structured matching.

\section{Experiments}

\subsection{Experimental Setup}

We evaluate 19 retrievers on ProSearch, spanning sparse lexical retrieval, general dense retrieval, scientific retrieval, and reasoning-oriented retrieval.
We include BM25~\citep{robertson2009probabilistic} as a classical sparse retriever; general-purpose dense retrievers~\citep{ma2023comprehensive} including Sentence-BERT~\citep{reimers2019sentence}, Harrier-OSS-v1-0.6B\footnote{\url{https://huggingface.co/microsoft/harrier-oss-v1-0.6b}}, BGE-M3~\citep{multi2024m3}, GTE-large~\citep{li2023towards}, NV-Embed-v2~\citep{lee2024nv}, and Qwen3-Embedding-8B~\citep{zhang2025qwen3}; scientific retrievers including SPECTER2~\citep{singh2023scirepeval}, SciNCL~\citep{ostendorff2022neighborhood}, PubMedBERT-base-embeddings~\citep{gu2021domain}, BioClinical-ModernBERT~\citep{sounack2025bioclinical}, ChemicalBERT-uncased~\footnote{\url{https://huggingface.co/recobo/chemical-bert-uncased}}, ChEmbed-prog~\citep{kasmaee2025chembed}, and MatSciBERT~\citep{gupta2022matscibert}; and reasoning-oriented retrievers including BGE-Reasoner-Embed~\citep{chen2025reasonembed}, ReasonIR-8B~\citep{shao2025reasonir}, ReasonEmbed~\citep{chen2025reasonembed}, INF-X-Retriever\footnote{\url{https://huggingface.co/infly/inf-retriever-v1-pro}}, and DIVER-Retriever~\citep{long2025diver}. We use nDCG@10 as the main evaluation metric. More details are provided in Appendix~\ref{app:experimental_detail}.

\subsection{Main Results}

\paragraph{ProSearch remains challenging for existing retrieval systems.} Results in Table~\ref{tab:retriever_results} show that ProSearch is highly challenging, with the best model achieving an average nDCG@10 of 0.290, indicating that existing retrieval models cannot reliably perform multi-constraint protocol retrieval. We observe that general-purpose dense retrievers achieve the best overall performance, which may be attributed to their broader training data coverage. In addition, scientific retrievers generally perform poorly, implying that adaptation on scientific text does not improve protocol retrieval performance. Interestingly, reasoning-oriented retrievers still lag behind the strongest general-purpose dense retrievers, suggesting that existing reasoning signals are not sufficient for ProSearch. The task requires not only domain-specific understanding, but also multi-constraint reasoning, including constraint dependencies within a coherent protocol. We also observe large performance differences across domains, with much lower scores on Immunotechnology and Electrochemistry, indicating that the difficulty of ProSearch is closely related to how experimental protocols are described and how constraints interact in different domains.
\begin{table*}[t]
\centering
\small
\setlength{\tabcolsep}{5pt}
\renewcommand{\arraystretch}{1.08}
\caption{
The performance of retrieval models on ProSearch.
We report nDCG@10 for six domains: Cell Culture (Cell.), Immunotechnology (Immuno.), Electrochemistry (Electro.), Organic Synthesis (Organic), Inorganic Materials (Inorg.), and Polymer Materials (Polymer).
Avg. denotes the average score across the six domains.
The best and second-best scores are bold and underlined, respectively.
}
\label{tab:retriever_results}
\begin{tabular*}{\textwidth}{@{\extracolsep{\fill}}lccccccc@{}}
\toprule
Retriever
& Cell.
& Immuno.
& Electro.
& Organic
& Inorg.
& Polymer
& Avg. \\
\midrule

\rowcolor{groupblue}
\multicolumn{8}{l}{Sparse lexical retrieval} \\
BM25
& 0.136 & 0.011 & 0.079 & 0.132 & 0.217 & 0.297 & 0.145 \\

\midrule
\rowcolor{groupblue}
\multicolumn{8}{l}{General-purpose dense retrieval} \\
Sentence-BERT
& 0.082 & 0.012 & 0.064 & 0.122 & 0.263 & 0.065 & 0.101 \\
GTE-large
& 0.222 & 0.041 & 0.127 & 0.184 & 0.441 & 0.182 & 0.199 \\
BGE-M3
& 0.254 & 0.025 & 0.106 & 0.216 & 0.369 & \textbf{0.326} & 0.216 \\
Harrier-OSS-v1-0.6B
& 0.261 & 0.049 & 0.159 & 0.198 & 0.467 & 0.177 & 0.219 \\
NV-Embed-v2
& \underline{0.358} & 0.070 & \textbf{0.184} & \textbf{0.290} & 0.448 & \underline{0.310} & \underline{0.277} \\
Qwen3-Embedding-8B
& \textbf{0.386} & \textbf{0.092} & \underline{0.175} & \underline{0.282} & \textbf{0.495} & 0.309 & \textbf{0.290} \\

\midrule
\rowcolor{groupblue}
\multicolumn{8}{l}{Scientific retrieval} \\
MatSciBERT
& 0.011 & 0.003 & 0.001 & 0.013 & 0.002 & 0.002 & 0.005 \\
ChemicalBERT-uncased
& 0.043 & 0.001 & 0.027 & 0.026 & 0.040 & 0.055 & 0.032 \\
SPECTER2
& 0.077 & 0.005 & 0.003 & 0.069 & 0.157 & 0.060 & 0.062 \\
SciNCL
& 0.098 & 0.034 & 0.036 & 0.099 & 0.190 & 0.020 & 0.080 \\
PubMedBERT-base-embeddings
& 0.069 & 0.001 & 0.039 & 0.104 & 0.213 & 0.059 & 0.081 \\
BioClinical-ModernBERT
& 0.142 & 0.011 & 0.053 & 0.093 & 0.272 & 0.084 & 0.109 \\
ChEmbed-prog
& 0.161 & 0.007 & 0.056 & 0.179 & 0.401 & 0.147 & 0.158 \\

\midrule
\rowcolor{groupblue}
\multicolumn{8}{l}{Reasoning-oriented retrieval} \\
ReasonIR-8B
& 0.079 & 0.006 & 0.021 & 0.120 & 0.092 & 0.043 & 0.060 \\
INF-X-Retriever
& 0.295 & 0.039 & 0.134 & 0.208 & 0.328 & 0.193 & 0.200 \\
ReasonEmbed
& 0.268 & 0.066 & 0.150 & 0.201 & 0.425 & 0.164 & 0.212 \\
BGE-Reasoner-Embed
& 0.285 & 0.069 & 0.152 & 0.240 & 0.435 & 0.215 & 0.233 \\
DIVER-Retriever
& 0.284 & \underline{0.072} & 0.170 & 0.262 & \underline{0.476} & 0.267 & 0.255 \\

\bottomrule
\end{tabular*}
\end{table*}
\begin{table*}[!t]
\centering
\small
\setlength{\tabcolsep}{4.5pt}
\renewcommand{\arraystretch}{1.08}
\caption{
The effect of query formulation strategies on ProSearch using Qwen3-Embedding-8B as the retriever.
We report nDCG@10 for six domains.
Avg. denotes the average score across the six domains, and $\Delta$Avg. denotes the change over the original query formulation.
The best score on each dataset is shown in bold and the second best is underlined.
}
\label{tab:query_formulation_qwen3}
\begin{tabular*}{\textwidth}{@{\extracolsep{\fill}}lcccccccc@{}}
\toprule
Query Formulation
& Cell.
& Immuno.
& Electro.
& Organic
& Inorg.
& Polymer
& Avg.
& $\Delta$Avg. \\
\midrule
Original
& \underline{0.386} & 0.092 & 0.175 & 0.282 & 0.495 & 0.309 & 0.290 & 0.000 \\
\midrule
Query Decomposition
& 0.317 & 0.068 & 0.091 & 0.205 & 0.454 & 0.254 & 0.232 & $-0.058$ \\
Structured Query
& 0.324 & 0.080 & 0.138 & 0.250 & 0.494 & 0.189 & 0.246 & $-0.044$ \\
INF-Query-Aligner\footnotemark
& 0.382 & 0.075 & 0.141 & 0.233 & 0.487 & 0.196 & 0.252 & $-0.038$ \\
HyDE~\citep{gao-etal-2023-precise}
& 0.359 & 0.098 & 0.154 & 0.262 & 0.501 & 0.255 & 0.272 & $-0.018$ \\
TongSearch-QR~\citep{Qin2025TongSearchQRRQ}
& 0.385 & 0.092 & 0.172 & 0.265 & \underline{0.505} & 0.288 & 0.285 & $-0.005$ \\
Reasoning-augmented~\citep{Su2024BRIGHTAR}
& \textbf{0.408} & \underline{0.111} & \underline{0.179} & \underline{0.309} & 0.494 & \textbf{0.359} & \underline{0.310} & \underline{$+0.020$} \\
DIVER-QExpand~\citep{long2025diver}
& 0.374 & \textbf{0.117} & \textbf{0.205} & \textbf{0.327} & \textbf{0.558} & \underline{0.318} & \textbf{0.317} & $\mathbf{+0.027}$ \\
\bottomrule
\end{tabular*}
\end{table*}

\xdef\ProSearchFootnoteNum{\number\value{footnote}}
\afterpage{%
  \footnotetext[\ProSearchFootnoteNum]{%
    \url{https://huggingface.co/infly/inf-query-aligner}%
  }%
}

\paragraph{Query rewriting brings limited gains.} Prior work has shown that query rewriting and query expansion can improve retrieval by reducing the mismatch between the original query and target documents~\citep{gao-etal-2023-precise, Jagerman2023QueryEB}. Therefore, we investigate whether query formulation affects protocol retrieval by comparing several variants using Qwen3-Embedding-8B (Table~\ref{tab:query_formulation_qwen3}). Detailed definitions and representative examples for each query variant are provided in Appendix~\ref{app:query_variants}. Results show that only DIVER-QExpand and Reasoning-augmented Query yield modest improvements over the original query. This indicates that while iterative feedback and explicit constraint emphasis are beneficial, the gains remain limited. Specifically, HyDE tends to introduce hallucinated details that dilute specific experimental parameters, while Query Decomposition fragments the protocol into partial matches, losing the necessary interdependencies between constraints. These error patterns suggest that a key bottleneck is retrievers' limited ability to jointly validate all constraints within a single experimental configuration.

\begin{table*}[t]
\centering
\small
\setlength{\tabcolsep}{5pt}
\renewcommand{\arraystretch}{1.08}
\caption{
The performance of reranking models on ProSearch.
We report nDCG@10 after reranking the top-100 candidate lists retrieved by Qwen3-Embedding-8B.
Avg. denotes the average score across the six domains.
The best score on each dataset is shown in bold and the second best is underlined.
}
\label{tab:reranker_results}
\begin{tabular*}{\textwidth}{@{\extracolsep{\fill}}lccccccc@{}}
\toprule
Model
& Cell.
& Immuno.
& Electro.
& Organic
& Inorg.
& Polymer
& Avg. \\
\midrule
\rowcolor{groupblue}
\multicolumn{8}{l}{First-stage retrieval} \\
Qwen3-Embedding-8B
& 0.386 & 0.092 & 0.175 & 0.282 & 0.495 & 0.309 & 0.290 \\

\midrule
\rowcolor{groupblue}
\multicolumn{8}{l}{Reranking on top-100 candidates} \\
BGE-Reranker-v2-m3~\citep{multi2024m3}
& 0.481 & 0.192 & 0.247 & 0.336 & \textbf{0.705} & 0.416 & 0.396 \\
Rank-R1~\citep{zhuang2025rank}
& 0.521 & 0.227 & 0.269 & 0.408 & 0.622 & 0.478 & 0.421 \\
Rank1~\citep{weller2025rank1}
& \textbf{0.592} & 0.266 & 0.241 & 0.416 & 0.598 & 0.457 & 0.428 \\
DIVER-GroupRank-7B~\citep{long2025grouprank}
& \underline{0.586} & 0.255 & 0.262 & 0.468 & \textbf{0.705} & \underline{0.531} & 0.468 \\
ReasonRank-8B~\citep{liu2025reasonrank}
& 0.577 & \underline{0.276} & \textbf{0.314} & \underline{0.483} & 0.637 & \textbf{0.547} & \underline{0.472} \\
Qwen3-Reranker~\citep{zhang2025qwen3}
& 0.579 & \textbf{0.301} & \underline{0.294} & \textbf{0.485} & \underline{0.692} & 0.490 & \textbf{0.474} \\
\bottomrule
\end{tabular*}
\end{table*}

\paragraph{Reranking substantially improves retrieval performance.} We further examine whether reranking can improve ProSearch by reranking the top-100 paragraphs retrieved by Qwen3-Embedding-8B with several representative rerankers. Results in Table~\ref{tab:reranker_results} show that reranking substantially improves retrieval performance, increasing the average nDCG@10 from 0.290 to 0.474. This indicates that the first-stage retriever can retrieve paragraphs containing relevant experimental constraints, but accurate ranking requires jointly judging multiple constraints, including their dependencies, and whether each constraint is fully satisfied. In addition, performance remains low on challenging domains such as Immunotechnology and Electrochemistry. Overall, reranking is effective but still insufficient, suggesting that significant room remains for improving fine-grained protocol ranking.

\paragraph{Expanding the retrieval scope makes protocol evidence harder to locate.}
\label{app:paragraph_pool_scope}
To examine retrieval performance beyond method sections, we expand the corpus to all body paragraphs while keeping the queries, relevance criteria, and paragraph-level retrieval unit unchanged. This increases the corpus size by 5.8$\times$ on average. As shown in Table~\ref{tab:paragraph_pool_scope}, all four representative retrievers experience performance drops, with Qwen3-Embedding-8B decreasing from 0.290 to 0.206 in average nDCG@10. However, their relative ranking remains unchanged. These results suggest that full-text paragraph retrieval introduces an additional evidence-localization challenge, while the overall retrieval conclusions remain consistent.

\begin{table*}[t]
\centering

\begin{minipage}[t]{0.53\textwidth}
\vspace{0pt}
\centering
\small
\setlength{\tabcolsep}{4pt}
\renewcommand{\arraystretch}{1.08}

\captionof{table}{
Comparison of retrieval performance over method paragraphs
and all body paragraphs across six domains, measured by
average nDCG@10.
}
\label{tab:paragraph_pool_scope}

\begin{tabular*}{\linewidth}{@{\extracolsep{\fill}}lcc@{}}
\toprule
Retriever & Method-only & All-body \\
\midrule
BM25                & 0.145 & 0.089 \\
BGE-Reasoner-Embed  & 0.233 & 0.164 \\
NV-Embed-v2         & 0.277 & 0.197 \\
Qwen3-Embedding-8B  & 0.290 & 0.206 \\
\bottomrule
\end{tabular*}
\end{minipage}
\hfill
\begin{minipage}[t]{0.45\textwidth}
\vspace{0pt}
\centering
\small
\setlength{\tabcolsep}{3pt}
\renewcommand{\arraystretch}{1.08}

\captionof{table}{
Open-world evidence retrieval performance on 30 ProSearch
queries, with five randomly sampled from each domain.
}
\label{tab:open_world_retrieval}

\begin{tabular*}{\linewidth}{@{\extracolsep{\fill}}lccc@{}}
\toprule
Method & Success@10 & P@10 & MRR \\
\midrule
Tavily Search & 0.300 & 0.077 & 0.209 \\
Tavily Agent  & 0.800 & 0.353 & 0.694 \\
\bottomrule
\end{tabular*}
\end{minipage}

\vspace{-2mm}
\end{table*}

\subsection{Analysis}

\paragraph{Explicit constraint matching provides a strong retrieval signal.} To assess the role of structured constraint information in
protocol retrieval, we compare text-based retrieval with two
diagnostic settings that use extracted protocol records.
As shown in Table~\ref{tab:structured_diagnostic},
Structured Diagnostic achieves an average nDCG@10 of 0.774,
compared with 0.290 for Qwen3-Embedding-8B, while
Hybrid Diagnostic further improves the score to 0.814.
These results suggest that matching constraints within
individual experimental configurations provides an effective
ranking signal that is not fully captured by text similarity.
These findings motivate retrieval approaches that
combine explicit constraint matching with semantic ranking.

\begin{table*}[t]
\centering
\small
\setlength{\tabcolsep}{5pt}
\renewcommand{\arraystretch}{1.08}

\caption{
Structured diagnostic analysis on average nDCG@10 across six domains.
Structured ranks paragraphs by the maximum number of query
constraints matched by any single extracted protocol record,
averaging over 20 randomized tie orderings.
Hybrid uses the same matching scores but breaks ties
using Qwen3-Embedding-8B similarity.
}
\label{tab:structured_diagnostic}

\begin{tabular*}{\textwidth}{
@{\extracolsep{\fill}}lrrrrrrr@{}
}
\toprule
Method
& Cell
& Electro
& Immuno
& Inorg
& Organic
& Polymer
& Avg. \\
\midrule
Qwen3-Embedding-8B
& 0.386 & 0.175 & 0.092 & 0.495 & 0.282 & 0.309 & 0.290 \\
Structured Diagnostic
& 0.789 & 0.698 & 0.682 & 0.875 & 0.773 & 0.829 & 0.774 \\
Hybrid Diagnostic
& 0.858 & 0.741 & 0.723 & 0.910 & 0.795 & 0.859 & 0.814 \\
\bottomrule
\end{tabular*}

\vspace{-2mm}
\end{table*}

\paragraph{Retrieval performance varies with query characteristics.}
We further analyze retrieval performance by constraint count,
numeric and negation constraints, and constraint roles.
As detailed in Appendix~\ref{query_characteristics},
constraint count is not a consistent indicator of retrieval
difficulty across domains.
Numeric constraints exhibit domain-dependent patterns,
while queries involving negation generally show lower
retrieval performance.
Role-level analysis further shows that queries containing
\textit{Condition} constraints achieve lower average nDCG@10
for both BM25 and Qwen3-Embedding-8B, whereas
\textit{Method} constraints show no comparable decrease.
Together, these findings highlight the importance of
interpreting experimental conditions in context and
associating them with the correct experimental configurations.

\paragraph{Agentic search improves open-world protocol retrieval.}
We further examine protocol retrieval in an open-world setting
using 30 queries, with five randomly sampled from each domain.
We compare two approaches: \textit{Tavily Search} uses the
Tavily API\footnote{\url{https://www.tavily.com/}} to search
the web with the original query and extract evidence passages,
while \textit{Tavily Agent} uses a LangChain
agent\footnote{\url{https://reference.langchain.com/python/langchain/agents/factory/create_agent}}
with Tavily tools to perform query reformulation and iterative search.
We evaluate Success@10, which measures whether at least one
relevant passage appears in the top 10; P@10, the proportion
of relevant passages in the top 10; and MRR, the mean reciprocal
rank of the first relevant passage.
As shown in Table~\ref{tab:open_world_retrieval},
Tavily Agent substantially outperforms direct web search
across all three metrics, demonstrating the benefit of
agentic retrieval for locating relevant experimental evidence.
Nevertheless, its P@10 remains at 0.353, indicating that
many retrieved passages still fail to satisfy all query constraints.

\begin{wrapfigure}{r}{0.5\textwidth}
\vspace{-2mm}
\centering
\includegraphics[width=\linewidth]{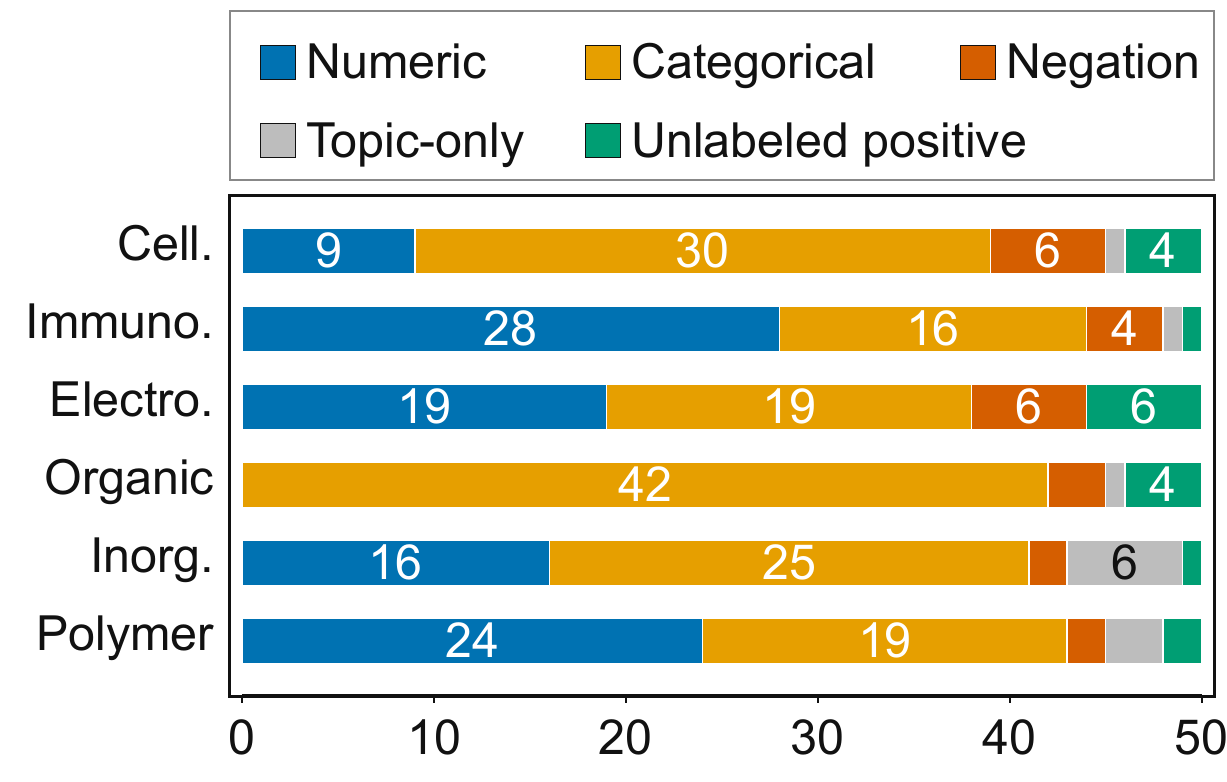}
\vspace{-1.0em}
\caption{
Manual error analysis of Qwen3-Embedding-8B. For each domain, we manually examine the two highest-ranked non-relevant results for each of 25 sampled queries.
}
\label{fig:error_breakdown}
\vspace{-2mm}
\end{wrapfigure}

\paragraph{Constraint violations dominate top-ranked retrieval errors.}
To characterize retrieval failures, we sample 25 queries
per domain and inspect the two highest-ranked paragraphs
labeled non-relevant by Qwen3-Embedding-8B for each query,
yielding 300 query--paragraph pairs.
Each pair is assigned a primary category: Numeric,
Categorical, or Negation violation, Topic-only, or
Unlabeled positive. When multiple constraints are violated,
we prioritize Negation over Numeric over Categorical. Representative examples are provided in Appendix~\ref{app:error_examples}.

As shown in Figure~\ref{fig:error_breakdown},
constraint violations account for 90\% of the examined cases,
with different error patterns across domains.
Categorical violations dominate in Cell Culture and
Organic Synthesis, whereas Numeric violations are especially
frequent in Immunotechnology and Polymer Materials.
In Immunotechnology, errors often involve missing quantitative
conditions or mismatched assay procedures, while
Electrochemistry exhibits mismatches in concentration,
scan parameters, and experimental systems.
These findings indicate that the main challenge is not
merely identifying relevant experimental topics, but
accurately determining whether the specified constraints
are satisfied within a single experimental configuration.

\section{Conclusion}

In this work, we introduce ProSearch, the first benchmark dedicated to multi-constraint protocol retrieval in experimental science. Our evaluation shows that current retrievers, including those specialized for reasoning, struggle to retrieve paragraphs that satisfy multiple experimental constraints within a single coherent protocol variant. Structured diagnostic analysis shows that explicit constraint matching provides a strong ranking signal, while manual error analysis highlights joint constraint binding as a key bottleneck. We hope ProSearch provides a new perspective for evaluating scientific retrieval and inspires further research into building more logically robust IR systems.

\section{Limitations}
\label{app:limitations}

While ProSearch provides a systematic evaluation of multi-constraint protocol retrieval, several limitations remain. First, the field schemas and constraint rules are designed by domain experts for selected experimental protocol types across six domains, limiting coverage of other experimental procedures and requiring additional expert effort for extension. Second, ProSearch focuses on individual method paragraphs and does not evaluate evidence integration across multiple paragraphs, tables, or supplementary materials. Third, LLM-based protocol extraction and relevance labeling, together with annotation pools constructed from extracted records, may introduce errors or omit relevant evidence. Although expert audits and robustness analyses support the stability of the main retrieval findings, systematic biases cannot be fully ruled out. Finally, while the queries are grounded in real experimental records, they are generated using templates and may not fully reflect researchers' actual information needs. Whether benchmark scores translate into retrieval usefulness in real scientific workflows also remains to be established. Future work will extend ProSearch to cross-paragraph protocol retrieval and incorporate researcher-authored queries and human evaluations to assess its practical applicability.

\section{Acknowledgments and Disclosures}

This work was supported by the Natural Science Foundation of China under Grant No. T2322027 and 62442204. We thank Ludi Wang, Yang Gao, Yang Chen, Wanping Xia, Shuaishuai Xu, Haolin Chen, and Xiaoyue Wang for their assistance with data annotation. 

The authors declare no competing interests.

\newpage
\appendix

\section{Broader Impacts}
\label{app:broader_impacts}

ProSearch is intended to support research on more reliable scientific evidence retrieval.
By focusing on method paragraphs and protocol-level constraints, it may help develop retrieval systems and scientific agents that better locate experimental evidence for protocol design, literature analysis, and reproducibility-oriented research.

However, ProSearch should not be treated as providing authoritative experimental recommendations.
Retrieval systems may still return incomplete or incorrect evidence, which could lead to flawed experimental decisions if used without expert verification.
The benchmark may also reflect biases in English open-access scientific literature.
Therefore, systems developed using ProSearch should be used as decision-support tools, and retrieved protocol evidence should be checked against the original papers before practical use.

\section{Details of Dataset Construction}
\label{app:dataset_construction}

\subsection{Expert-Guided Protocol Modeling}
\label{app:expert_guided_protocol_modeling}

\paragraph{Field schemas.}
Each domain is associated with an expert-defined field schema that specifies the protocol fields used for query construction.
Each field is annotated with its role, value type, and whether it can be used in negated constraints or as an exploration field.
Tables~\ref{tab:schema_cell_culture}--\ref{tab:schema_polymer_materials} list the complete field schemas for the six domains.

\begin{table*}[ht]
\centering
\scriptsize
\setlength{\tabcolsep}{4pt}
\renewcommand{\arraystretch}{1.18}
\renewcommand{\tabularxcolumn}[1]{m{#1}}
\caption{
Field schema used to represent experimental protocols in the cell culture domain.
Neg. and Exp. indicate whether the field can be used in negated constraints and exploration queries, respectively, where T denotes true and F denotes false.
}
\label{tab:schema_cell_culture}
\begin{tabularx}{\textwidth}{
>{\raggedright\arraybackslash}m{0.28\textwidth}
>{\raggedright\arraybackslash}m{0.10\textwidth}
>{\raggedright\arraybackslash}m{0.13\textwidth}
>{\centering\arraybackslash}m{0.05\textwidth}
>{\centering\arraybackslash}m{0.05\textwidth}
>{\raggedright\arraybackslash}X
}
\toprule
Field & Role & Value Type & Neg. & Exp. & Description \\
\midrule

\makecell[l]{\texttt{cultured\_cells.}\\\texttt{cell\_type}}
& Goal
& Categorical
& F
& F
& Cell line or cell type, such as A549, HeLa, or primary fibroblasts.
\\
\rowrule

\makecell[l]{\texttt{culture\_medium.}\\\texttt{basal\_medium}}
& Input
& Categorical
& T
& T
& Basal culture medium, such as DMEM, RPMI-1640, or IMDM.
\\
\rowrule

\makecell[l]{\texttt{culture\_medium.serum.}\\\texttt{identity}}
& Input
& Categorical
& T
& T
& Serum identity or serum-free condition, such as fetal bovine serum, horse serum, or serum-free medium.
\\
\rowrule

\texttt{culture\_substrate\_or\_coating}
& Input
& Categorical
& T
& T
& Culture substrate or coating, such as collagen-coated plates, Matrigel, or poly-D-lysine.
\\
\rowrule

\makecell[l]{\texttt{culture\_medium.serum.}\\\texttt{concentration}}
& Condition
& Numeric / Range
& F
& T
& Serum concentration, usually expressed as a percentage.
\\
\rowrule

\texttt{seeding\_density}
& Condition
& Numeric / Range
& F
& T
& Cell seeding density, usually expressed as cells per well, cells per dish, or cells per area.
\\
\rowrule

\texttt{differentiation\_state}
& Condition
& Categorical
& F
& F
& Cell differentiation or activation state; retained in extraction but not used for query construction.
\\
\rowrule

\texttt{confluence}
& Condition
& Numeric / Range
& F
& T
& Cell confluence before treatment or passage, usually expressed as a percentage.
\\
\rowrule

\texttt{culture\_mode}
& Method
& Categorical
& F
& T
& Culture mode, such as 2D monolayer culture, 3D spheroid culture, or suspension culture.
\\

\bottomrule
\end{tabularx}
\end{table*}

\paragraph{Field dependency rules.}
Some fields require additional contextual fields to form meaningful query constraints.
For example, a seeding density value should be interpreted together with the corresponding cell type and culture mode, while an electrochemical potential should be anchored by the reference electrode and measurement method.
Table~\ref{tab:dependency_rules} lists the complete dependency rules used during query construction.

\paragraph{Role-level templates.}
Role-level templates define abstract query structures using protocol roles.
Each template consists of a fixed role template and, when applicable, an exploration role.
The complete set of role-level templates is shown in Table~\ref{tab:role_level_templates}.

\paragraph{Field-level templates.}
Field-level templates instantiate the role-level templates with concrete fields from each domain schema.
We enumerate valid field combinations and filter them using field dependency rules and field-level usage flags.
Table~\ref{tab:field_level_template_stats} reports the resulting template statistics, and Table~\ref{tab:field_level_template_examples} gives representative examples.

\begin{table*}[ht]
\centering
\scriptsize
\setlength{\tabcolsep}{4pt}
\renewcommand{\arraystretch}{1.18}
\renewcommand{\tabularxcolumn}[1]{m{#1}}
\caption{
Field schema for the immunotechnology domain.
}
\label{tab:schema_immunotechnology}
\begin{tabularx}{\textwidth}{
>{\raggedright\arraybackslash}m{0.28\textwidth}
>{\raggedright\arraybackslash}m{0.10\textwidth}
>{\raggedright\arraybackslash}m{0.13\textwidth}
>{\centering\arraybackslash}m{0.05\textwidth}
>{\centering\arraybackslash}m{0.05\textwidth}
>{\raggedright\arraybackslash}X
}
\toprule
Field & Role & Value Type & Neg. & Exp. & Description \\
\midrule

\makecell[l]{\texttt{protein\_targets.}\\\texttt{target}}
& Goal
& Categorical
& F
& F
& Protein target measured in the assay, such as AKT, ERK, or GAPDH.
\\
\rowrule

\texttt{normalization\_control}
& Input
& Categorical
& T
& T
& Normalization or loading control used for comparison, such as $\beta$-actin, GAPDH, or tubulin.
\\
\rowrule

\texttt{sample\_type}
& Input
& Categorical
& F
& F
& Type of biological sample, such as cell lysate, tissue lysate, serum, or purified protein.
\\
\rowrule

\makecell[l]{\texttt{blocking\_solution.}\\\texttt{identity}}
& Input
& Categorical
& T
& T
& Identity of the blocking solution, such as milk, BSA, or skim milk.
\\
\rowrule

\makecell[l]{\texttt{protein\_targets.}\\\texttt{target\_form}}
& Condition
& Categorical
& F
& F
& Target form or modification state, such as phosphorylated or total protein; retained in extraction but not used for query construction.
\\
\rowrule

\makecell[l]{\texttt{blocking\_solution.}\\\texttt{concentration}}
& Condition
& Numeric / Range
& F
& T
& Concentration of the blocking solution, typically expressed as a percentage.
\\
\rowrule

\texttt{gel\_percentage}
& Condition
& Numeric / Range
& F
& T
& SDS-PAGE gel percentage; retained as an exploration field but not used as a fixed query constraint.
\\
\rowrule

\texttt{loading\_amount}
& Condition
& Numeric / Range
& F
& T
& Sample loading amount, typically expressed in units such as $\mu$g per lane.
\\
\rowrule

\texttt{lysis\_buffer}
& Method
& Categorical
& F
& T
& Lysis buffer used for protein extraction, such as RIPA buffer or NP-40 lysis buffer.
\\

\bottomrule
\end{tabularx}
\end{table*}
\begin{table*}[!ht]
\centering
\scriptsize
\setlength{\tabcolsep}{4pt}
\renewcommand{\arraystretch}{1.18}
\renewcommand{\tabularxcolumn}[1]{m{#1}}
\caption{
Field schema for the electrochemistry domain.
}
\label{tab:schema_electrochemistry}
\begin{tabularx}{\textwidth}{
>{\raggedright\arraybackslash}m{0.28\textwidth}
>{\raggedright\arraybackslash}m{0.10\textwidth}
>{\raggedright\arraybackslash}m{0.13\textwidth}
>{\centering\arraybackslash}m{0.05\textwidth}
>{\centering\arraybackslash}m{0.05\textwidth}
>{\raggedright\arraybackslash}X
}
\toprule
Field & Role & Value Type & Neg. & Exp. & Description \\
\midrule

\texttt{working\_electrode}
& Method
& Categorical
& F
& F
& Working electrode material or type, such as glassy carbon, platinum, or nickel foam.
\\
\rowrule

\makecell[l]{\texttt{electrolytes.}\\\texttt{chemical}}
& Input
& Categorical
& T
& T
& Electrolyte chemical species, such as KOH, H$_2$SO$_4$, Na$_2$SO$_4$, or KCl.
\\
\rowrule

\texttt{reference\_electrode}
& Method
& Categorical
& F
& T
& Reference electrode used in the electrochemical setup, such as Ag/AgCl, SCE, or RHE.
\\
\rowrule

\texttt{target\_species}
& Goal
& Categorical
& F
& F
& Electrochemical target species or reaction subject, such as oxygen, hydrogen, nitrate, or glucose.
\\
\rowrule

\makecell[l]{\texttt{electrolytes.}\\\texttt{concentration}}
& Condition
& Numeric / Range
& F
& T
& Electrolyte concentration, typically expressed in units such as M, mM, or wt\%.
\\
\rowrule

\texttt{temperature}
& Condition
& Numeric / Range
& F
& T
& Experimental temperature, usually expressed in $^\circ$C.
\\
\rowrule

\texttt{scan\_rate}
& Condition
& Numeric / Range
& F
& T
& Potential scan rate used in voltammetry measurements, typically expressed in mV/s.
\\
\rowrule

\texttt{applied\_potential}
& Condition
& Numeric / Range
& F
& T
& Applied or scanned potential, typically expressed in V; its interpretation usually depends on the reference electrode and measurement method.
\\
\rowrule

\texttt{measurement\_method}
& Method
& Categorical
& F
& T
& Electrochemical measurement method, such as cyclic voltammetry, linear sweep voltammetry, chronoamperometry, or EIS.
\\

\bottomrule
\end{tabularx}
\end{table*}
\begin{table*}[ht]
\centering
\scriptsize
\setlength{\tabcolsep}{4pt}
\renewcommand{\arraystretch}{1.18}
\renewcommand{\tabularxcolumn}[1]{m{#1}}
\caption{
Field schema for the organic synthesis domain.
}
\label{tab:schema_organic_synthesis}
\begin{tabularx}{\textwidth}{
>{\raggedright\arraybackslash}m{0.28\textwidth}
>{\raggedright\arraybackslash}m{0.10\textwidth}
>{\raggedright\arraybackslash}m{0.13\textwidth}
>{\centering\arraybackslash}m{0.05\textwidth}
>{\centering\arraybackslash}m{0.05\textwidth}
>{\raggedright\arraybackslash}X
}
\toprule
Field & Role & Value Type & Neg. & Exp. & Description \\
\midrule

\texttt{transformation\_type}
& Goal
& Categorical
& F
& F
& Type of chemical transformation, such as cross-coupling, arylation, amination, borylation, or C--H activation.
\\
\rowrule

\texttt{substrates}
& Input
& Categorical
& F
& T
& Reactant or substrate class used in the reaction, such as aryl halides, boronic acids, amines, or alkynes.
\\
\rowrule

\makecell[l]{\texttt{catalysts.}\\\texttt{name}}
& Input
& Categorical
& F
& T
& Catalyst identity or catalytic species, such as Pd, Ni, Cu catalysts, or named catalyst complexes.
\\
\rowrule

\texttt{ligands}
& Input
& Categorical
& F
& T
& Ligand used with the catalyst system, such as phosphine, bipyridine, or N-heterocyclic carbene ligands.
\\
\rowrule

\makecell[l]{\texttt{catalysts.}\\\texttt{loading}}
& Condition
& Numeric
& F
& T
& Catalyst loading, typically expressed as mol\%; retained as an exploration field but not used as a fixed query constraint.
\\
\rowrule

\texttt{bases}
& Condition
& Categorical
& T
& T
& Base used in the reaction, such as K$_2$CO$_3$, Cs$_2$CO$_3$, NaOtBu, or triethylamine.
\\
\rowrule

\texttt{additives}
& Condition
& Categorical
& F
& T
& Additive or auxiliary reagent used to promote or modify the reaction.
\\
\rowrule

\texttt{solvents}
& Condition
& Categorical
& T
& T
& Reaction solvent or solvent mixture, such as toluene, dioxane, DMF, DMSO, or water.
\\
\rowrule

\texttt{activation\_mode}
& Method
& Categorical
& F
& T
& Activation or reaction mode, such as thermal heating, photoredox activation, electrochemical activation, or microwave irradiation.
\\

\bottomrule
\end{tabularx}
\end{table*}
\begin{table*}[ht]
\centering
\scriptsize
\setlength{\tabcolsep}{4pt}
\renewcommand{\arraystretch}{1.18}
\renewcommand{\tabularxcolumn}[1]{m{#1}}
\caption{
Field schema for the inorganic materials domain.
}
\label{tab:schema_inorganic_materials}
\begin{tabularx}{\textwidth}{
>{\raggedright\arraybackslash}m{0.28\textwidth}
>{\raggedright\arraybackslash}m{0.10\textwidth}
>{\raggedright\arraybackslash}m{0.13\textwidth}
>{\centering\arraybackslash}m{0.05\textwidth}
>{\centering\arraybackslash}m{0.05\textwidth}
>{\raggedright\arraybackslash}X
}
\toprule
Field & Role & Value Type & Neg. & Exp. & Description \\
\midrule

\texttt{target\_material}
& Goal
& Categorical
& F
& F
& Target inorganic material or composition to be synthesized, such as an oxide, perovskite, spinel, or ceramic material.
\\
\rowrule

\texttt{precursors}
& Input
& Categorical
& F
& T
& Starting precursor materials used for synthesis, such as metal oxides, carbonates, nitrates, or salts.
\\
\rowrule

\makecell[l]{\texttt{dopant\_or\_additive.}\\\texttt{identity}}
& Input
& Categorical
& T
& T
& Identity of the dopant or additive introduced into the material system.
\\
\rowrule

\makecell[l]{\texttt{dopant\_or\_additive.}\\\texttt{amount}}
& Condition
& Numeric / Range
& F
& T
& Amount or concentration of the dopant or additive, such as molar ratio, atomic percentage, or weight percentage.
\\
\rowrule

\texttt{atmosphere}
& Condition
& Categorical
& T
& T
& Processing atmosphere, such as air, oxygen, nitrogen, argon, vacuum, or reducing atmosphere.
\\
\rowrule

\texttt{heating\_rate}
& Condition
& Numeric / Range
& F
& T
& Heating rate used during thermal processing, usually expressed in $^\circ$C/min.
\\
\rowrule

\texttt{pressing\_pressure}
& Condition
& Numeric / Range
& F
& T
& Pressure used for pellet pressing or compaction, usually expressed in MPa.
\\
\rowrule

\makecell[l]{\texttt{sintering.}\\\texttt{temperature}}
& Condition
& Numeric / Range
& F
& T
& Sintering temperature, usually expressed in $^\circ$C.
\\
\rowrule

\makecell[l]{\texttt{sintering.}\\\texttt{time}}
& Condition
& Numeric / Range
& F
& T
& Sintering duration, usually expressed in hours.
\\
\rowrule

\makecell[l]{\texttt{post\_treatment.}\\\texttt{temperature}}
& Condition
& Numeric / Range
& F
& T
& Temperature used in post-treatment steps such as annealing, calcination, or heat treatment.
\\
\rowrule

\makecell[l]{\texttt{post\_treatment.}\\\texttt{time}}
& Condition
& Numeric / Range
& F
& T
& Duration of post-treatment steps such as annealing, calcination, or heat treatment.
\\
\rowrule

\texttt{synthesis\_route}
& Method
& Categorical
& F
& T
& Synthesis route or processing method, such as solid-state reaction, sol-gel synthesis, or ceramic processing.
\\
\rowrule

\makecell[l]{\texttt{post\_treatment.}\\\texttt{type}}
& Method
& Categorical
& F
& T
& Type of post-treatment operation, such as annealing, calcination, sintering, quenching, or heat treatment.
\\

\bottomrule
\end{tabularx}
\end{table*}
\begin{table*}[t]
\centering
\scriptsize
\setlength{\tabcolsep}{4pt}
\renewcommand{\arraystretch}{1.18}
\renewcommand{\tabularxcolumn}[1]{m{#1}}
\caption{
Field schema for the polymer materials domain.
}
\label{tab:schema_polymer_materials}
\begin{tabularx}{\textwidth}{
>{\raggedright\arraybackslash}m{0.28\textwidth}
>{\raggedright\arraybackslash}m{0.10\textwidth}
>{\raggedright\arraybackslash}m{0.13\textwidth}
>{\centering\arraybackslash}m{0.05\textwidth}
>{\centering\arraybackslash}m{0.05\textwidth}
>{\raggedright\arraybackslash}X
}
\toprule
Field & Role & Value Type & Neg. & Exp. & Description \\
\midrule

\texttt{polymer\_system}
& Goal
& Categorical
& F
& F
& Polymer or resin system under study, such as PEGDA, GelMA, polyurethane acrylate, or epoxy acrylate.
\\
\rowrule

\texttt{polymer\_concentration}
& Condition
& Numeric
& F
& T
& Polymer concentration in the formulation, typically expressed as wt\%, vol\%, or concentration per unit volume.
\\
\rowrule

\makecell[l]{\texttt{photoinitiators.}\\\texttt{name}}
& Input
& Categorical
& T
& T
& Identity of the photoinitiator, such as Irgacure 2959, LAP, or TPO.
\\
\rowrule

\makecell[l]{\texttt{crosslinking\_agent.}\\\texttt{name}}
& Input
& Categorical
& T
& T
& Identity of the crosslinking agent, such as MBAA, PEGDA, or other crosslinking reagents.
\\
\rowrule

\makecell[l]{\texttt{photoinitiators.}\\\texttt{loading}}
& Condition
& Numeric
& F
& T
& Amount or loading of the photoinitiator, typically expressed as wt\%, mol\%, or phr.
\\
\rowrule

\makecell[l]{\texttt{crosslinking\_agent.}\\\texttt{loading}}
& Condition
& Numeric
& F
& T
& Amount or loading of the crosslinking agent, typically expressed as wt\%, mol\%, or phr.
\\
\rowrule

\texttt{wavelength}
& Condition
& Numeric / Range
& F
& T
& Irradiation wavelength used for curing or polymerization, usually expressed in nm.
\\
\rowrule

\texttt{intensity}
& Condition
& Numeric / Range
& F
& T
& Light intensity during irradiation, usually expressed in mW/cm$^2$.
\\
\rowrule

\texttt{exposure\_time}
& Condition
& Numeric / Range
& F
& T
& Irradiation or exposure time, usually expressed in minutes or seconds.
\\
\rowrule

\texttt{light\_source\_type}
& Method
& Categorical
& F
& T
& Type of light source used for curing or photopolymerization, such as UV lamp, UV LED, mercury lamp, or visible-light source.
\\

\bottomrule
\end{tabularx}
\end{table*}
\begin{table*}[ht]
\centering
\small
\setlength{\tabcolsep}{4pt}
\renewcommand{\arraystretch}{1.08}
\caption{
Field-level template statistics by domain.
No Exp. denotes templates without an exploration field.
Input, Cond., and Method denote templates whose exploration field belongs to the corresponding protocol role.
The columns 3, 4, and 5 denote the total number of constraints in a template.
Numeric denotes templates containing at least one numeric field, Neg. denotes templates containing at least one field eligible for negation, and Neither denotes templates containing neither numeric nor negation-eligible fields.
}
\label{tab:field_level_template_stats}
\begin{tabular}{lrrrrrrrrrrr}
\toprule
\multirow{2}{*}{Domain}
& \multicolumn{4}{c}{Explore Role}
& \multicolumn{3}{c}{\# Constraints}
& \multicolumn{3}{c}{Template Properties}
& \multirow{2}{*}{Total} \\
\cmidrule(lr){2-5}
\cmidrule(lr){6-8}
\cmidrule(lr){9-11}
& No Exp. & Input & Cond. & Method
& 3 & 4 & 5
& Numeric & Neg. & Neither
& \\
\midrule
Cell Culture        & 9   & 13 & 10  & 5  & 9  & 15  & 13  & 28  & 31  & 0  & 37  \\
Electrochemistry    & 17  & 5  & 25  & 8  & 16 & 24  & 15  & 50  & 44  & 0  & 55  \\
Immunotechnology    & 7   & 6  & 9   & 7  & 10 & 13  & 6   & 23  & 12  & 4  & 29  \\
Inorganic Materials & 103 & 65 & 198 & 42 & 55 & 136 & 217 & 361 & 251 & 10 & 408 \\
Organic Synthesis   & 23  & 26 & 47  & 15 & 31 & 50  & 30  & 8   & 60  & 47 & 111 \\
Polymer Materials   & 22  & 16 & 31  & 13 & 20 & 37  & 25  & 75  & 76  & 0  & 82  \\
\bottomrule
\end{tabular}
\end{table*}

\begin{table*}[ht]
\centering
\small
\setlength{\tabcolsep}{6pt}
\renewcommand{\arraystretch}{1.15}
\caption{
Field dependency rules used during query construction.
Each rule specifies the contextual fields that must be known when the trigger field is used.
}
\label{tab:dependency_rules}
\begin{tabularx}{\textwidth}{
>{\raggedright\arraybackslash}p{0.43\textwidth}
>{\raggedright\arraybackslash}X
}
\toprule
Trigger Field & Required Fields \\
\midrule

\multicolumn{2}{l}{\textbf{Cell Culture}} \\
\rowrule
\texttt{culture\_medium.serum.concentration}
& \texttt{culture\_medium.serum.identity}, \texttt{culture\_medium.basal\_medium} \\
\texttt{culture\_medium.serum.identity}
& \texttt{culture\_medium.basal\_medium} \\
\texttt{seeding\_density}
& \texttt{cultured\_cells.cell\_type}, \texttt{culture\_mode} \\
\texttt{differentiation\_state}
& \texttt{cultured\_cells.cell\_type} \\
\texttt{confluence}
& \texttt{cultured\_cells.cell\_type}, \texttt{culture\_mode} \\

\midrule
\multicolumn{2}{l}{\textbf{Immunotechnology}} \\
\rowrule
\texttt{protein\_targets.target\_form}
& \texttt{protein\_targets.target} \\
\texttt{blocking\_solution.concentration}
& \texttt{blocking\_solution.identity} \\
\texttt{gel\_percentage}
& \texttt{protein\_targets.target} \\
\texttt{normalization\_control}
& \texttt{sample\_type} \\
\texttt{loading\_amount}
& \texttt{sample\_type} \\
\texttt{lysis\_buffer}
& \texttt{sample\_type} \\

\midrule
\multicolumn{2}{l}{\textbf{Electrochemistry}} \\
\rowrule
\texttt{electrolytes.concentration}
& \texttt{electrolytes.chemical} \\
\texttt{applied\_potential}
& \texttt{reference\_electrode}, \texttt{measurement\_method} \\
\texttt{scan\_rate}
& \texttt{measurement\_method} \\

\midrule
\multicolumn{2}{l}{\textbf{Organic Synthesis}} \\
\rowrule
\texttt{catalysts.loading}
& \texttt{catalysts.name} \\
\texttt{ligands}
& \texttt{catalysts.name} \\

\midrule
\multicolumn{2}{l}{\textbf{Inorganic Materials}} \\
\rowrule
\texttt{dopant\_or\_additive.amount}
& \texttt{dopant\_or\_additive.identity} \\
\texttt{post\_treatment.temperature}
& \texttt{post\_treatment.type} \\
\texttt{post\_treatment.time}
& \texttt{post\_treatment.type} \\
\texttt{pressing\_pressure}
& \texttt{synthesis\_route} \\

\midrule
\multicolumn{2}{l}{\textbf{Polymer Materials}} \\
\rowrule
\texttt{photoinitiators.loading}
& \texttt{photoinitiators.name} \\
\texttt{crosslinking\_agent.loading}
& \texttt{crosslinking\_agent.name} \\
\texttt{wavelength}
& \texttt{photoinitiators.name} \\
\texttt{intensity}
& \texttt{photoinitiators.name}, \texttt{wavelength} \\
\texttt{exposure\_time}
& \texttt{photoinitiators.name}, \texttt{wavelength} \\

\bottomrule
\end{tabularx}
\end{table*}
\begin{table*}[!t]
\centering
\small
\setlength{\tabcolsep}{4pt}
\renewcommand{\arraystretch}{1.08}
\caption{
Examples of protocol fields and constraints in Cell Culture.
Protocol fields are organized by their roles and instantiated with operators and values to form query constraints.
Dependent fields denote expert-defined dependencies used to avoid underspecified constraints, such as requiring seeding density to be anchored by the corresponding cell type.
}
\label{tab:protocol_schema_examples}
\resizebox{\textwidth}{!}{
\begin{tabular}{cccccc}
\toprule
Role & Field & Operator & Value & Explanation & Dependent Fields \\
\midrule
Goal
& Cell type
& $=$
& HUVEC
& Uses HUVEC cells
& -- \\

Input
& Basal medium
& $\neq$
& DMEM
& Does not use DMEM
& -- \\

Condition
& Seeding density
& $=$
& $6{\times}10^4$ cells/well
& Seeded at $6{\times}10^4$ cells/well
& Cell type \\

Condition
& Seeding density
& \textsc{between}
& $500$--$1{,}000$ cells/well
& Seeded within this density range
& Cell type \\

Method
& Culture mode
& $=$
& 3D
& Uses 3D culture
& -- \\
\bottomrule
\end{tabular}}
\end{table*}

\begin{table*}[ht]
\centering
\small
\setlength{\tabcolsep}{4pt}
\renewcommand{\arraystretch}{1.12}
\caption{
Role-level templates used for query construction.
Fixed Role Template specifies the protocol roles instantiated as fixed query constraints.
Explore Role denotes the role of the field whose value is to be recovered; ``--'' indicates that the template does not include an exploration field.
The number of constraints counts both fixed constraints and the exploration field when present.
}
\label{tab:role_level_templates}
\begin{tabularx}{\textwidth}{
>{\centering\arraybackslash}p{0.14\textwidth}
>{\raggedright\arraybackslash}X
>{\centering\arraybackslash}p{0.18\textwidth}
>{\centering\arraybackslash}p{0.16\textwidth}
}
\toprule
Template ID & Fixed Role Template & Explore Role & \# Constraints \\
\midrule

T1  & Goal + Input + Condition & -- & 3 \\
T2  & Goal + Input + Method & -- & 3 \\
T3  & Goal + Input + Input + Condition & -- & 4 \\
T4  & Goal + Input + Condition + Method & -- & 4 \\
T5  & Goal + Input + Condition + Condition + Method & -- & 5 \\

\midrule

T6  & Goal + Input & Input & 3 \\
T7  & Goal + Condition & Input & 3 \\
T8  & Goal + Input + Condition & Input & 4 \\
T9  & Goal + Condition + Method & Input & 4 \\
T10 & Goal + Input + Condition + Method & Input & 5 \\

\midrule

T11 & Goal + Input & Condition & 3 \\
T12 & Goal + Method & Condition & 3 \\
T13 & Goal + Input + Condition & Condition & 4 \\
T14 & Goal + Input + Method & Condition & 4 \\
T15 & Goal + Input + Condition + Method & Condition & 5 \\

\midrule

T16 & Goal + Input & Method & 3 \\
T17 & Goal + Condition & Method & 3 \\
T18 & Goal + Input + Condition & Method & 4 \\
T19 & Goal + Input + Input & Method & 4 \\
T20 & Goal + Input + Input + Condition & Method & 5 \\

\bottomrule
\end{tabularx}
\end{table*}
\begin{table*}[p]
\centering
\small
\setlength{\tabcolsep}{4pt}
\renewcommand{\arraystretch}{1.16}
\caption{
Representative field-level templates in the inorganic materials domain.
Each example is shown in two rows: the role-level template specifies the fixed roles and exploration role, while the field-level template instantiates them with concrete fields from the domain schema.
}
\label{tab:field_level_template_examples}
\begin{tabularx}{\textwidth}{
>{\centering\arraybackslash}p{0.12\textwidth}
>{\raggedright\arraybackslash}p{0.16\textwidth}
>{\raggedright\arraybackslash}X
>{\raggedright\arraybackslash}p{0.20\textwidth}
}
\toprule
Template ID & Level & Fixed Template & Explore Role/Field \\
\midrule

\multirow{2}{*}{T1}
& Role-level
& Goal + Input + Condition
& -- \\
& Field-level
& \texttt{target\_material} + \texttt{dopant\_or\_additive.identity} + \texttt{atmosphere}
& -- \\

\midrule

\multirow{2}{*}{T5}
& Role-level
& Goal + Input + Condition + Condition + Method
& -- \\
& Field-level
& \texttt{target\_material} + \texttt{precursors} + \texttt{sintering.temperature} + \texttt{sintering.time} + \texttt{synthesis\_route}
& -- \\

\midrule

\multirow{2}{*}{T6}
& Role-level
& Goal + Input
& Input \\
& Field-level
& \texttt{target\_material} + \texttt{dopant\_or\_additive.identity}
& \texttt{precursors} \\

\midrule

\multirow{2}{*}{T14}
& Role-level
& Goal + Input + Method
& Condition \\
& Field-level
& \texttt{target\_material} + \texttt{precursors} + \texttt{synthesis\_route}
& \texttt{sintering.time} \\

\midrule

\multirow{2}{*}{T16}
& Role-level
& Goal + Input
& Method \\
& Field-level
& \texttt{target\_material} + \texttt{dopant\_or\_additive.identity}
& \texttt{post\_treatment.type} \\

\midrule

\multirow{2}{*}{T20}
& Role-level
& Goal + Input + Input + Condition
& Method \\
& Field-level
& \texttt{target\_material} + \texttt{dopant\_or\_additive.identity} + \texttt{precursors} + \texttt{sintering.time}
& \texttt{synthesis\_route} \\

\bottomrule
\end{tabularx}
\end{table*}

\FloatBarrier

\subsection{Corpus Collection and Paragraph Filtering}
\label{app:corpus_filtering}

\paragraph{Article retrieval.}
We collect open-access articles from OpenAlex using domain-specific search queries designed by domain experts.
The queries aim to retrieve experimental articles that are likely to contain method-level details, while excluding reviews, surveys, perspectives, and other non-experimental article types.
Table~\ref{tab:search_queries} lists the complete search queries used for the six experimental domains.

\begin{table*}[ht]
\centering
\scriptsize
\setlength{\tabcolsep}{5pt}
\renewcommand{\arraystretch}{1.12}
\renewcommand{\tabularxcolumn}[1]{m{#1}}
\caption{
Search queries used to collect candidate articles for each experimental domain.
}
\label{tab:search_queries}
\begin{tabularx}{\textwidth}{
>{\raggedright\arraybackslash}p{0.14\textwidth}
>{\raggedright\arraybackslash}X
}
\toprule
Domain & Search Query \\
\midrule

Cell Culture
&
(cell culture OR cell line OR in vitro)
AND (DMEM OR RPMI OR IMDM OR DMEM/F12 OR DMEM-F12 OR alpha-MEM OR McCoy's 5A OR FBS OR fetal bovine serum OR serum-free OR incubat* OR seeded OR seeding density OR confluence OR culture medium)
NOT (review OR meta-analysis OR survey OR perspective)
\\

\midrule

Immunotechnology
&
(western blot OR immunoblot OR immunoblotting OR western blotting)
AND (primary antibody OR secondary antibody OR membrane OR SDS-PAGE OR 1: OR ECL OR TBST)
NOT (review OR meta-analysis OR survey OR perspective)
\\

\midrule

Electrochemistry
&
(electrochem* OR electrochemical OR electrocatal* OR cyclic voltammetry OR linear sweep voltammetry OR voltammetry OR potentiostat OR galvanostat)
AND (electrolyte OR electrode OR three-electrode OR reference electrode OR Ag/AgCl OR RHE)
NOT (review OR meta-analysis OR survey OR perspective OR benchmarking OR standardization OR protocol)
\\

\midrule

Organic Synthesis
&
(cross-coupling OR cross coupling OR coupling OR arylation OR amination OR borylation OR Suzuki OR Heck OR Sonogashira OR Buchwald-Hartwig OR C-H activation OR photoredox OR electrochemical)
AND (metal-catalyzed OR transition-metal-catalyzed OR catalyzed OR catalytic OR catalyst OR Pd OR Ni OR Cu)
AND (ligand OR base OR solvent OR reaction conditions OR atmosphere)
NOT (review OR meta-analysis OR survey OR perspective OR editorial)
\\

\midrule

Inorganic Materials
&
(solid-state synthesis OR solid state synthesis OR solid-state reaction OR solid state reaction OR conventional solid-state OR ceramic method OR mixed-oxide OR mixed oxide OR calcination OR sintering OR anneal* OR heat treatment OR heat-treat* OR firing OR fired)
AND (ceramic* OR oxide* OR metal oxide* OR perovskite* OR spinel* OR ferrite* OR titanate*)
NOT (review OR meta-analysis OR survey OR perspective)
\\

\midrule

Polymer Materials
&
(photopolymer* OR photopolymerization OR UV cur* OR UV-cur* OR UV curable OR photo cur* OR light cur* OR UV irradiation OR light irradiation OR photoinitiator OR photo-initiator OR UV lamp OR UV LED OR UV-LED)
NOT (review OR meta-analysis OR survey OR perspective)
\\

\bottomrule
\end{tabularx}
\end{table*}

\paragraph{Method paragraph pool construction.}
After retrieving candidate articles, we download the available full texts and parse them with GROBID into TEI XML files. We remove paragraphs from footnotes, endnotes, reference lists, bibliographies, captions, figures, and tables.
To construct the method paragraph pool, we identify method-related sections from the parsed top-level section headings.
A section is selected if its heading matches case-insensitive heading patterns covering both generic experimental sections and domain-specific method sections.
Table~\ref{tab:method_section_keywords} summarizes the main heading patterns used for method section identification. Within the selected method-related sections, we extract TEI paragraph elements as candidate method paragraphs.
We also remove paragraphs longer than 2048 GPT-2 tokens.
Each retained method paragraph is stored with its article DOI, section heading, paragraph text, and source type.

\begin{table*}[t]
\centering
\scriptsize
\setlength{\tabcolsep}{5pt}
\renewcommand{\arraystretch}{1.12}
\renewcommand{\tabularxcolumn}[1]{m{#1}}
\caption{
Section heading patterns used to identify method-related sections from GROBID TEI outputs.
The patterns include generic experimental headings and domain-specific headings for the six experimental domains.
}
\label{tab:method_section_keywords}
\begin{tabularx}{\textwidth}{
>{\raggedright\arraybackslash}p{0.18\textwidth}
>{\raggedright\arraybackslash}X
}
\toprule
Category & Heading Patterns \\
\midrule

Generic method headings
&
experimental section, experimental part, experimental details, experimental methods, experimental procedures, experimental setup, experimental design, experimental conditions, experiments, materials and methods, methods and materials, methods, methodology, materials preparation, sample preparation, preparation of samples, preparation of materials, preparation of compounds, preparation of polymers, synthesis and characterization, characterization, characterization methods, measurements, instrumentation, analytical methods
\\

\midrule

Cell culture
&
cell culture, cell culture methods, cell culture conditions, culture conditions, cell lines, in vitro experiments, cell treatment, cell treatments, incubation conditions
\\

\midrule

Immunotechnology
&
western blot, western blotting, immunoblot, immunoblotting, ELISA, flow cytometry, immunofluorescence, immunostaining, immunohistochemistry, antibodies, SDS-PAGE, protein extraction, sample processing
\\

\midrule

Electrochemistry
&
electrochemical methods, electrochemical measurements, electrochemical characterization, electrochemical testing, electrochemical evaluation, electrode preparation, electrolyte preparation, cell assembly, electrochemical setup, cyclic voltammetry, linear sweep voltammetry, chronoamperometry, chronopotentiometry, impedance spectroscopy, EIS
\\

\midrule

Organic synthesis
&
general procedure, typical procedure, synthetic procedure, synthetic details, synthesis of, preparation of, reaction conditions, catalytic conditions, substrate scope, optimization of reaction, optimization of conditions
\\

\midrule

Inorganic materials
&
solid-state synthesis, solid state synthesis, solid-state reaction, solid state reaction, calcination, sintering, sintered, annealing, annealed, firing, pellet preparation, powder preparation, ceramic processing, fabrication, materials fabrication
\\

\midrule

Polymer Materials
&
photopolymerization, photopolymerization procedure, UV curing, UV-curing, photocuring, photo-curing, light curing, UV irradiation, light irradiation, photoinitiated polymerization, photoinitiator, resin preparation, formulation preparation, monomer formulation, sample preparation, film preparation, hydrogel preparation, curing conditions
\\

\bottomrule
\end{tabularx}
\end{table*}

\paragraph{Full paragraph pool construction.}
In addition to the method paragraph pool, we construct a full paragraph pool for the full-text retrieval analysis.
This pool is extracted from all TEI body sections rather than only matched method-related sections.
We apply the same paragraph-level filters to remove footnotes, endnotes, reference lists, bibliographies, captions, figures, tables, and length outliers.

\paragraph{Corpus construction statistics.}
Table~\ref{tab:corpus_construction_stats} reports the corpus construction statistics for each domain, including the number of retrieved articles, downloaded full texts, method paragraph candidates, and full paragraph candidates.

\begin{table}[t]
\centering
\small
\setlength{\tabcolsep}{5pt}
\renewcommand{\arraystretch}{1.08}
\caption{
Corpus construction statistics by domain.
Retrieved Articles denotes the number of records returned by the OpenAlex search queries.
Downloaded Articles denotes the number of full texts successfully downloaded.
Method Pool Size denotes the number of paragraphs extracted from matched method-related sections.
Full Pool Size denotes the number of retained body paragraphs used for full-text retrieval analysis.
}
\label{tab:corpus_construction_stats}
\begin{tabular*}{\textwidth}{@{\extracolsep{\fill}}lcccc@{}}
\toprule
Domain & Retrieved Articles & Downloaded Articles & Method Pool Size & Full Pool Size \\
\midrule
Cell Culture        & 15,600 & 9,633  & 46,004 & 311,042 \\
Immunotechnology    & 33,400 & 19,909 & 93,315 & 644,077 \\
Electrochemistry    & 42,040 & 13,879 & 91,375 & 409,133 \\
Organic Synthesis   & 14,420 & 7,669  & 30,371 & 197,506 \\
Inorganic Materials & 26,910 & 8,794  & 46,136 & 240,380 \\
Polymer Materials   & 18,045 & 9,360  & 50,977 & 283,907 \\
\bottomrule
\end{tabular*}
\end{table}

\FloatBarrier

\subsection{Query Construction Details}
\label{app:query_generation_details}

This section provides additional details on query construction in ProSearch.
We describe how candidate constraint sets are instantiated from protocol records, how they are selected and sampled, how weak constraint sets are filtered, and how the retained constraint sets are converted into natural-language queries.

\paragraph{Instantiating constraint sets from protocol records.}
Given the field-level templates described in Section~\ref{app:expert_guided_protocol_modeling}, we instantiate candidate constraint sets using the normalized protocol records.
For each field-level template, all fields are filled using values from a single protocol record.
If any required field is missing in that record, the candidate is discarded.
This prevents the resulting constraint set from combining conditions that come from different experimental configurations.
Categorical values are instantiated as equality constraints, numeric point values as equality constraints, and numeric intervals as range constraints.
The allowed operators for each field follow the field schema.

\paragraph{Selecting and sampling constraint sets.}
For each candidate constraint set, we compute its support count, defined as the number of unique paragraphs containing at least one protocol record that satisfies all constraints.
We retain candidates with support counts between 3 and 8.
The lower bound ensures that each retained candidate has multiple structurally supported paragraphs before validation, while the upper bound removes constraint sets that are too frequent to represent a specific protocol search need.
For candidate constraint sets with an exploration field, we additionally require the explored field to have multiple distinct observed values among the supporting records.
This avoids generating questions whose answer is either unavailable or nearly constant in the corpus.
For constraint sets with negation, we only use negated values that occur sufficiently often in the corpus, so that the negated condition reflects a meaningful alternative rather than a rare value.
From the remaining candidates, we sample constraint sets for each domain.
Sampling is stratified by constraint count, so that the final benchmark includes queries with different levels of constraint complexity.

\paragraph{Filtering constraint sets with an LLM.}
We further use GPT-5.4 to assess whether each sampled constraint set forms a meaningful protocol retrieval request.
This filtering step targets cases that satisfy the automatic construction rules but remain scientifically underspecified or too broad for retrieval evaluation.
Common rejected cases include constraint sets with missing experimental context, overly broad negated constraints, or explored fields whose values depend on omitted setup information.
Examples of rejected constraint sets and their rejection reasons are shown in Table~\ref{tab:rejected_constraint_examples}.

\begin{table*}[t]
\centering
\footnotesize
\setlength{\tabcolsep}{5pt}
\renewcommand{\arraystretch}{1.12}
\renewcommand{\tabularxcolumn}[1]{m{#1}}
\caption{
Representative constraint sets rejected during LLM-based usability filtering.
These cases are structurally valid but are excluded since they do not define well-specified or scientifically meaningful protocol retrieval requests.
}
\label{tab:rejected_constraint_examples}
\begin{tabularx}{\textwidth}{@{}p{0.64\textwidth}X@{}}
\toprule
Rejected constraint set & Rejection reason \\
\midrule

\textbf{Cell culture}

\begin{tabular}{@{}lll@{}}
\texttt{culture\_medium.basal\_medium} & $=$ & \texttt{RPMI-1640} \\
\texttt{culture\_medium.serum.concentration} & $=$ & \texttt{10\%} \\
\texttt{culture\_medium.serum.identity} & $=$ & \texttt{fetal bovine serum} \\
\texttt{cultured\_cells.cell\_type} & $=$ & \texttt{prostate cancer} \\
\end{tabular}
&
The cell type value is too broad to define a stable cell culture target.
\\

\midrule

\textbf{Electrochemistry}

\begin{tabular}{@{}lll@{}}
\texttt{electrolytes.chemical} & $\neq$ & \texttt{PBS} \\
\texttt{measurement\_method} & $=$ & \texttt{EIS} \\
\texttt{scan\_rate} & $=$ & \texttt{50 mV/s} \\
\texttt{target\_species} & $=$ & \texttt{K$_3$[Fe(CN)$_6$]/K$_4$[Fe(CN)$_6$]} \\
\end{tabular}
&
The setup is methodologically inconsistent because scan rate is not a stable defining parameter for EIS.
\\

\midrule

\textbf{Immunotechnology}

\begin{tabular}{@{}lll@{}}
\texttt{blocking\_solution.concentration} & $=$ & \texttt{5\%} \\
\texttt{blocking\_solution.identity} & $=$ & \texttt{skim milk} \\
\texttt{protein\_targets.target} & $=$ & \texttt{LIMK1} \\
\end{tabular}
&
The constraint set is dominated by a minor blocking condition and lacks stronger assay anchors such as sample type or normalization context.
\\

\midrule

\textbf{Inorganic materials}

\begin{tabular}{@{}lll@{}}
\texttt{precursors} & $=$ & \texttt{alumina} \\
\texttt{sintering.temperature} & $=$ & \texttt{1600~$^\circ$C} \\
\texttt{target\_material} & $=$ & \texttt{Al$_2$O$_3$} \\
\end{tabular}
&
The precursor and target material describe the same alumina system, making the constraint set redundant and weak as a retrieval request.
\\

\midrule

\textbf{Organic synthesis}

\begin{tabular}{@{}lll@{}}
\texttt{solvents} & $\neq$ & \texttt{DME} \\
\texttt{substrates} & $=$ & \texttt{N-(prop-2-yn-1-yl)benzamide} \\
\texttt{transformation\_type} & $=$ & \texttt{Suzuki coupling} \\
\end{tabular}
&
The substrate and transformation type do not form a coherent reaction setup, so the solvent query is not well anchored.
\\

\midrule

\textbf{Polymer materials}

\begin{tabular}{@{}lll@{}}
\texttt{photoinitiators.name} & $=$ & \texttt{dye} \\
\texttt{polymer\_system} & $=$ & \texttt{PEGDA} \\
\texttt{wavelength} & $=$ & \texttt{405 nm} \\
\end{tabular}
&
The photoinitiator value is too generic to define a reliable PEGDA photocuring setup.
\\

\bottomrule
\end{tabularx}
\end{table*}

\paragraph{Generating natural-language queries.}
After constraint set filtering, GPT-5.4 is used to convert each retained constraint set into a natural-language query.
The generated queries are then filtered by Claude Sonnet 4.6 for faithfulness, clarity, and retrieval usefulness.
Rejected examples and their rejection reasons are provided in Table~\ref{tab:rejected_query_examples}.

\begin{table*}[t]
\centering
\footnotesize
\setlength{\tabcolsep}{5pt}
\renewcommand{\arraystretch}{1.08}

\caption{
Representative generated queries rejected during LLM-based query quality filtering.
Each example is generated from a structured constraint set but fails at least one of the three quality checks: faithfulness (F), clarity (C), and retrieval usefulness (U).
}
\label{tab:rejected_query_examples}
\begin{tabularx}{\textwidth}{@{}p{0.56\textwidth}X@{}}
\toprule
Rejected query and structured constraints & Scores and rejection reason \\
\midrule

\textbf{Cell culture}

\textit{Query:} What seeding density is used for A549 cells cultured in 2D?

\textit{Structured constraints:} \par\smallskip
\begin{tabular}{@{}lll@{}}
\texttt{culture\_mode} & $=$ & \texttt{2D} \\
\texttt{cultured\_cells.cell\_type} & $=$ & \texttt{A549} \\
\texttt{seeding\_density} & $\neq$ & \texttt{null} \\
\end{tabular}
&
\textit{Scores:} F=1, C=1, U=0.

\textit{Reason:}
The fixed context is too sparse. A549 cells in 2D culture define a common baseline setup, so the query is too generic to retrieve specific experimental evidence.
\\

\midrule

\textbf{Immunotechnology}

\textit{Query:} What normalization control other than AKT is used for PTEN Western blotting in tissue lysate with 30 $\mu$g loading?

\textit{Structured constraints:} \par\smallskip
\begin{tabular}{@{}lll@{}}
\texttt{loading\_amount} & $=$ & \texttt{30 $\mu$g} \\
\texttt{normalization\_control} & $\neq$ & \texttt{AKT} \\
\texttt{protein\_targets.target} & $=$ & \texttt{PTEN} \\
\texttt{sample\_type} & $=$ & \texttt{tissue lysate} \\
\end{tabular}
&
\textit{Scores:} F=1, C=1, U=0.

\textit{Reason:}
AKT is a signaling protein rather than a standard normalization control, so the negated normalization-control constraint is scientifically unstable.
\\

\midrule

\textbf{Organic synthesis}

\textit{Query:} What solvent is used for cross-coupling of aryl halides?

\textit{Structured constraints:} \par\smallskip
\begin{tabular}{@{}lll@{}}
\texttt{substrates} & $=$ & \texttt{aryl halide} \\
\texttt{transformation\_type} & $=$ & \texttt{cross-coupling} \\
\texttt{solvents} & $\neq$ & \texttt{null} \\
\end{tabular}
&
\textit{Scores:} F=1, C=1, U=0.

\textit{Reason:}
The fixed conditions are too broad. Aryl-halide cross-coupling covers many reaction classes, catalysts, and bases, so the solvent target is not meaningfully constrained.
\\

\midrule

\textbf{Polymer materials}

\textit{Query:} Find photocuring experiments using TMPTA with DQ1 as the photoinitiator at 1.35$\times$10$^{-3}$--1.80$\times$10$^{-3}$ M.

\textit{Structured constraints:} \par\smallskip
\begin{tabular}{@{}lll@{}}
\texttt{photoinitiators.loading} & \texttt{between} & \texttt{1.35--1.80$\times$10$^{-3}$ M} \\
\texttt{photoinitiators.name} & $=$ & \texttt{DQ1} \\
\texttt{polymer\_system} & $=$ & \texttt{TMPTA} \\
\end{tabular}
&
\textit{Scores:} F=1, C=1, U=0.

\textit{Reason:}
The molarity-based loading range is an unstable retrieval anchor for this polymer curing setup, where photoinitiator loading is more commonly specified by mass or molar fraction.
\\

\bottomrule
\end{tabularx}
\end{table*}

\paragraph{Construction and filtering statistics.}
Table~\ref{tab:constraint_set_construction_stats} summarizes the main statistics of constraint set construction and query filtering.
The construction statistics show that the template-based procedure produces a large pool of candidate constraint sets in each domain.
Automatic filtering and sampling then reduce this pool to a controlled set of candidates for LLM-based usability checking.
This indicates that query construction in ProSearch is not based on arbitrary field combinations, but is constrained by structured support, field usage rules, and sampling requirements.

\begin{table*}[t]
\centering
\small
\setlength{\tabcolsep}{5pt}
\renewcommand{\arraystretch}{1.12}
\caption{
Constraint set construction and filtering statistics.
Field-level templates denote the number of field-level templates used for each domain.
Constructed sets are candidate constraint sets instantiated from field-level templates and protocol records.
After automatic filtering denotes deduplicated candidates after support, exploration-field, and negation-related filtering.
Eligible for sampling denotes candidates satisfying the final sampling requirements.
Sampled sets are selected for LLM-based usability filtering.
}
\label{tab:constraint_set_construction_stats}
\begin{tabular*}{\textwidth}{@{\extracolsep{\fill}}lcccccc@{}}
\toprule
Statistic 
& Cell. & Electro. & Immuno. & Inorg. & Organic & Polymer \\
\midrule
Field-level templates      & 37     & 55      & 29      & 408     & 111     & 82 \\
Constructed sets           & 16,181 & 143,381 & 434,912 & 648,472 & 171,753 & 23,177 \\
After automatic filtering  & 723    & 10,018  & 38,232  & 1,906   & 10,271  & 876 \\
Eligible for sampling      & 723    & 7,083   & 38,232  & 1,906   & 10,271  & 876 \\
Sampled sets               & 300    & 300     & 300     & 300     & 300     & 300 \\
Final evaluation queries   & 284    & 201     & 264     & 247     & 252     & 267 \\
\bottomrule
\end{tabular*}
\end{table*}

\begin{figure*}[t]
    \centering
    \includegraphics[width=\textwidth]{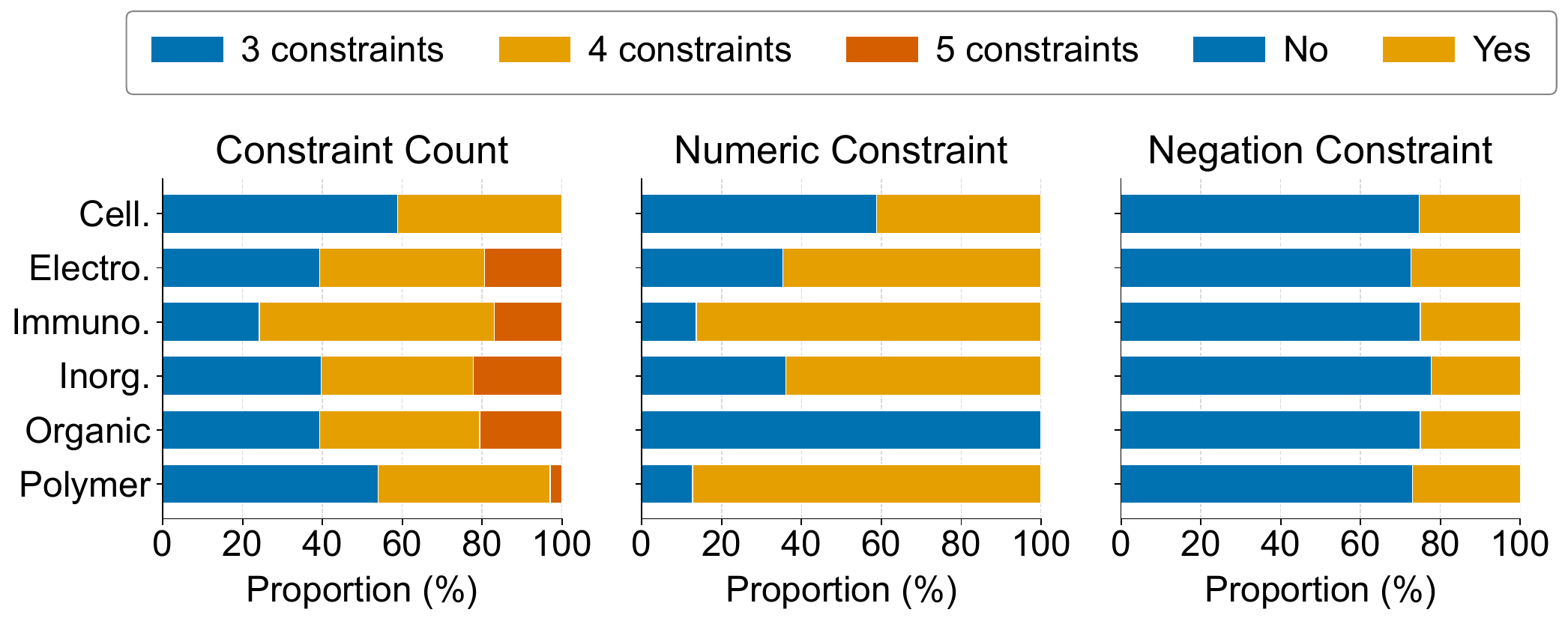}
    \caption{
    Query composition of ProSearch across domains.
    The three panels show the distributions of constraint counts, numeric constraints, and negation constraints, respectively.
    }
    \label{fig:query_composition}
\end{figure*}

\paragraph{Constraint diversity.}
Figure~\ref{fig:query_composition} characterizes the constraint patterns of the final query set after construction and filtering.
Across domains, ProSearch is dominated by 3- and 4-constraint queries, while 5-constraint queries are retained when sufficient structured support is available.
This distribution reflects the realism constraint in our query design: the queries are more specific than broad topical searches, but avoid overly narrow combinations that domain experts would be unlikely to ask in practice.
Numeric constraints are frequent in strongly parameterized domains such as electrochemistry, inorganic materials, and polymer materials.
Negation constraints appear in a substantial fraction of queries, allowing ProSearch to test whether retrievers can exclude protocols with disallowed conditions rather than only match positive evidence.
Overall, these statistics show that ProSearch queries are grounded in real structured records while covering diverse constraint patterns beyond topical matching.

\FloatBarrier

\section{Experimental Details}

\subsection{Models and instructions}
\label{app:models_instructions}

For each retrieval model evaluated in ProSearch, Table~\ref{tab:model_details} provides the model category, parameter size, architecture, official maximum input length for queries and documents, whether instructions are used, the specific checkpoint, and the license.
\begin{table*}[t]
\centering
\small
\setlength{\tabcolsep}{5pt}
\renewcommand{\arraystretch}{1.1}
\caption{
Details of the 19 retrieval models evaluated in \textsc{ProSearch}.
We report the model size, architecture, official maximum input length, whether an instruction is used for retrieval, the checkpoint or implementation, and the license.
Max $|Q|$ and Max $|D|$ denote the official maximum input length for queries and documents, respectively.
For BM25, the model size is not applicable and the maximum input length is unbounded in principle.
For models whose license is not explicitly stated in the public model card, we report the most likely license based on the upstream release.
}
\label{tab:model_details}
\resizebox{\textwidth}{!}{
\begin{tabular}{l c c c c c l l} 
\toprule
Model & Size & Architecture & Max $|Q|$ & Max $|D|$ & Instruction & Version / Checkpoint & License \\
\midrule

\multicolumn{8}{l}{\textit{Sparse lexical retrieval}} \\
BM25 & N/A & Sparse & $\infty$ & $\infty$ & No & rank\_bm25.BM25Okapi & Apache-2.0 \\

\midrule
\multicolumn{8}{l}{\textit{General-purpose dense retrieval}} \\
Sentence-BERT & 109M & Encoder & 384 & 384 & No & sentence-transformers/all-mpnet-base-v2 & Apache-2.0 \\
Harrier-OSS-v1 & 0.6B & Decoder & 32768 & 32768 & No & microsoft/harrier-oss-v1-0.6b & MIT \\
BGE-M3 & 568M & Encoder & 8192 & 8192 & No & BAAI/bge-m3 & MIT \\
GTE-large & 434M & Encoder & 8192 & 8192 & No & Alibaba-NLP/gte-large-en-v1.5 & Apache-2.0 \\
NV-Embed-v2 & 8B & Decoder & 32768 & 32768 & Yes & nvidia/NV-Embed-v2 & CC-BY-NC-4.0 \\
Qwen3-Embedding-8B & 8B & Decoder & 32768 & 32768 & Yes & Qwen/Qwen3-Embedding-8B & Apache-2.0 \\

\midrule
\multicolumn{8}{l}{\textit{Scientific retrieval}} \\
SPECTER2 & 110M & Encoder & 512 & 512 & No & allenai/specter2\_base & Apache-2.0 \\
SciNCL & 110M & Encoder & 512 & 512 & No & malteos/scincl & MIT \\
PubMedBERT-base & 109M & Encoder & 512 & 512 & No & NeuML/pubmedbert-base-embeddings & Apache-2.0 \\
BioClinical-ModernBERT & 150M & Encoder & 8192 & 8192 & No & thomas-sounack/BioClinical-ModernBERT-base & MIT \\
ChemicalBERT-uncased & 110M & Encoder & 512 & 512 & No & recobo/chemical-bert-uncased & Apache-2.0 \\
ChEmbed-prog & 137M & Encoder & 8192 & 8192 & No & BASF-AI/ChEmbed-prog & CC-BY-NC-4.0 \\
MatSciBERT & 110M & Encoder & 512 & 512 & No & m3rg-iitd/matscibert & MIT \\

\midrule
\multicolumn{8}{l}{\textit{Reasoning-oriented retrieval}} \\
BGE-Reasoner-Embed & 8B & Decoder & 32768 & 32768 & Yes & BAAI/bge-reasoner-embed-qwen3-8b-0923 & Apache-2.0 \\
ReasonIR-8B & 8B & Decoder & 131072 & 131072 & No & reasonir/ReasonIR-8B & CC-BY-NC-4.0 \\
ReasonEmbed & 8B & Decoder & 32768 & 32768 & Yes & hanhainebula/reason-embed-qwen3-8b-0928 & Apache-2.0 \\
INF-X-Retriever & 7B & Decoder & 32768 & 32768 & Yes & infly/inf-retriever-v1-pro & Apache-2.0 \\
DIVER-Retriever & 4B & Decoder & 40000 & 40000 & No & AQ-MedAI/Diver-Retriever-4B & Apache-2.0 \\

\bottomrule
\end{tabular}
}
\end{table*}

Table~\ref{tab:model_instructions} lists the instructions and adapter settings used in our experiments.
Instruction-based retrievers prepend a task description to the query following the format used in our implementation.
Models without instructions encode the original query and paragraph text directly.

\begin{table*}[t]
\centering
\small
\setlength{\tabcolsep}{5pt}
\renewcommand{\arraystretch}{1.12}
\caption{
Instructions used for retrieval models in ProSearch.
The placeholder \{query\} denotes the original natural-language query.
}
\label{tab:model_instructions}
\begin{tabularx}{\textwidth}{p{0.40\textwidth}X}
\toprule
Models & Query input  \\
\midrule

NV-Embed-v2, BGE-Reasoner-Embed, INF-X-Retriever, ReasonEmbed
&
Instruct: Given a web search query, retrieve paragraphs that answer the question

Query: \{query\}

\\

\midrule
Qwen3-Embedding-8B, Harrier-OSS-v1, DIVER-Retriever
&
Instruct: Given a question, retrieve paragraphs that answer the question

Query:\{query\}
\\

\bottomrule
\end{tabularx}
\end{table*}

\subsection{Query variant experiments}
\label{app:query_variants}

To evaluate whether transforming the original natural-language query improves protocol retrieval, we generate several query variants for each query.
These variants are used as retrieval inputs while keeping the document corpus, retriever, and evaluation protocol unchanged.
This setting allows us to isolate the effect of query formulation from changes in the retrieval model.

\begin{table*}[p]
\centering
\footnotesize
\setlength{\tabcolsep}{0pt}
\renewcommand{\arraystretch}{1.12}
\caption{
Prompts used for query variant experiments.
}
\label{tab:query_variant_prompts}
\begin{tabularx}{\textwidth}{@{}X@{}}
\toprule

\textbf{\textit{Query Decomposition}} \\
\midrule
\begin{minipage}{0.98\linewidth}
Given a retrieval query, decompose it into a small set of simpler retrieval sub-queries.

First, identify internally the query intent and the constraints in the query.
Then generate the smallest non-redundant set of sub-queries that together cover the original constraints.

Requirements:
(1) Each sub-query must remain a natural question.
(2) Each sub-query must preserve the original query intent.
For an exploration query, it must still ask about the same field as the original query.
For a reproduction query, it must still ask for a setup/protocol/paragraph satisfying constraints.
(3) Each sub-query should keep only part of the original constraints, so that it is simpler than the original query.
(4) Different sub-queries should reflect different retained constraints, not paraphrases of the same reduced query.
(5) Together, the sub-queries should cover all original constraints.
(6) Use as few sub-queries as possible.
(7) Do not introduce new constraints or change the meaning of the original query.

Examples:

Query: What applied potential is used (vs SCE) at 25$^\circ$C with an Au working electrode?

Decomposition:
[
``What applied potential is used (vs SCE) with an Au working electrode?'',
``What applied potential is used (vs SCE) at 25$^\circ$C?''
]

Query: Find an electrochemistry experimental protocol describing cyclic voltammetry (CV) performed in acetate buffer using an AGCE working electrode.

Decomposition:
[
``Find an electrochemistry experimental protocol describing cyclic voltammetry (CV) performed in acetate buffer.'',
``Find an electrochemistry experimental protocol describing cyclic voltammetry (CV) using an AGCE working electrode.''
]

Return only a JSON-style list of plain question strings.

Query: \{query\}
\end{minipage}
\\

\midrule
\textbf{\textit{Structured Query}} \\
\midrule
\begin{minipage}{0.98\linewidth}
No generation prompt is used.
We directly use the structured constraint representation as the retrieval input.
\end{minipage}
\\

\midrule
\textbf{\textit{HyDE}} \\
\midrule
\begin{minipage}{0.98\linewidth}
Given the query, write a short scientific paragraph that would likely answer it.
Preserve the key experimental constraints.

Query: \{query\}
\end{minipage}
\\

\midrule
\textbf{\textit{Reasoning-augmented Query}} \\
\midrule
\begin{minipage}{0.98\linewidth}
Given the query, generate a retrieval-oriented reasoning process for experimental setup retrieval.

The reasoning should briefly explain what kind of experimental setup, condition combination, or explored field is being sought, what information a relevant paragraph would explicitly contain, and which constraints must be preserved exactly.
Focus on how the requested conditions would be expressed together in a relevant experimental description.

Return only the reasoning process.

Query: \{query\}
\end{minipage}
\\

\midrule
\textbf{\textit{DIVER-QExpand}} \\
\midrule
\begin{minipage}{0.98\linewidth}
\textit{Initial round.}

Given a query and the provided passages, most of which may be incorrect or irrelevant, identify helpful information from the passages and use it to write a correct answering passage.
Use your own knowledge, not just the example passages!

Query: \{query\}

Possible helpful passages: \{top\_passages\_str\}

\vspace{0.5em}
\textit{Refinement round.}

Given a query, the provided passages, most of which may be incorrect or irrelevant, and the previous round's answer, identify helpful information from the passages and refine the prior answer.
Ensure the output directly addresses the original query.
Use your own knowledge, not just the example passages!

Query: \{query\}

Possible helpful passages: \{top\_passages\_str\}

Prior generated answer: \{last\_round\_answer\}
\end{minipage}
\\

\midrule
\textbf{\textit{INF-Query-Aligner}} \\
\midrule
\begin{minipage}{0.98\linewidth}
For the input query, formulating a concise search query for dense retrieval by distilling the core intent from a complex user prompt and ignoring LLM instructions.
The response should be less than 200 words.
\end{minipage}
\\

\midrule
\textbf{\textit{TongSearch-QR}} \\
\midrule
\begin{minipage}{0.98\linewidth}
Instructions:

Rewrite the following user query into a search query for retrieval.

Preserve all important constraints faithfully, including entities, conditions, negation, numbers, and ranges.

Do not add new constraints.
Do not remove important constraints.

Return only the rewritten query, without explanation.

Query: \{query\}
\end{minipage}
\\

\bottomrule
\end{tabularx}
\end{table*}

We consider six types of query variants:
\begin{itemize}
    \item \textbf{Query Decomposition} splits the original query into a set of non-redundant sub-queries, each emphasizing a different part of the protocol constraints. The retrieval results from all sub-queries are then merged into a single ranked list.

    \item \textbf{Structured Query} uses the structured constraint representation directly as the retrieval input.

    \item \textbf{HyDE} first generates a hypothetical method paragraph that could answer the query and then uses this generated paragraph as the retrieval input.

    \item \textbf{Reasoning-augmented Query} augments the original query with retrieval-oriented reasoning, including the target protocol, expected evidence, and constraints that should be preserved.

    \item \textbf{DIVER-QExpand} performs iterative query expansion by incorporating explicit reasoning and retrieval feedback.

    \item \textbf{Model-based query rewriting} rewrites the original query using existing query rewriting models, including INF-Query-Aligner and TongSearch-QR.
\end{itemize}

For HyDE, Query Decomposition, Reasoning-augmented Query, and DIVER-QExpand, we use GPT-5.4 to generate the query variants.
For INF-Query-Aligner and TongSearch-QR, we use the released rewriting models directly.
Table~\ref{tab:query_variant_prompts} summarizes the prompts or input settings used for different query variants.
Table~\ref{tab:query_variant_examples} provides examples of the resulting query variants.

\begin{table*}[t]
\centering
\footnotesize
\setlength{\tabcolsep}{0pt}
\renewcommand{\arraystretch}{1.12}
\caption{
Examples of query variants generated from the same original ProSearch query. The DIVER-QExpand example is shortened for readability.
}
\label{tab:query_variant_examples}
\begin{tabularx}{\textwidth}{@{}X@{}}
\toprule

\textbf{\textit{Original Query}} \\
\midrule
Find 3D culture experiments in which C2C12 cells are grown in alginate.
\\

\midrule
\textbf{\textit{Query Decomposition}} \\
\midrule
\begin{minipage}{0.98\linewidth}
[
``Find 3D culture experiments in which C2C12 cells are grown.'',

``Find 3D culture experiments in which cells are grown in alginate.''
]
\end{minipage}
\\

\midrule
\textbf{\textit{Structured Query}} \\
\midrule
task=reproduction; culture\_mode=3D; culture\_substrate\_or\_coating=alginate; cultured\_cells.cell\_type=C2C12
\\

\midrule
\textbf{\textit{HyDE}} \\
\midrule
\begin{minipage}{0.98\linewidth}
Several 3D culture studies have embedded C2C12 murine myoblasts in alginate hydrogels under controlled crosslinking and differentiation conditions.
Typical experiments suspend C2C12 cells in sterile sodium alginate, form beads or bulk gels by ionic crosslinking with CaCl$_2$, and culture them in growth medium before switching to low-serum differentiation medium to assess myogenesis in 3D.
In these systems, investigators often vary alginate concentration, stiffness, bead size, or incorporation of adhesive motifs such as RGD to test effects on viability, proliferation, alignment, myotube formation, and expression of myogenic markers including MyoD, myogenin, and myosin heavy chain.
Key constraints common to these experiments are the use of C2C12 cells specifically in 3D alginate matrices rather than 2D coatings, calcium-crosslinked gel formation, and evaluation of differentiation over several days to weeks.
\end{minipage}
\\

\midrule
\textbf{\textit{Reasoning-augmented Query}} \\
\midrule
\begin{minipage}{0.98\linewidth}
We need paragraphs describing a 3D cell culture experimental setup where the cell type is C2C12 and the matrix or material is alginate.
A relevant paragraph would explicitly state that C2C12 cells were cultured, encapsulated, or embedded in alginate in a 3D construct such as alginate beads, hydrogels, capsules, scaffolds, or microfibers.
The key constraints that must be preserved exactly are the C2C12 cell line, the 3D culture setting, and the use of alginate as the culture matrix.
Paragraphs that mention C2C12 and alginate separately without making clear that the cells are cultured within a 3D alginate system should be deprioritized.
\end{minipage}
\\

\midrule
\textbf{\textit{DIVER-QExpand}} \\
\midrule
\begin{minipage}{0.98\linewidth}
3D culture experiments where C2C12 myoblasts are grown in alginate include alginate microcapsule or bead cultures, where C2C12 cells are encapsulated in sodium alginate and crosslinked with CaCl$_2$ to form a 3D culture system.
Relevant passages may describe cells resuspended in 1.5\% sodium alginate, dripped into CaCl$_2$ to form capsules, and cultured in wells for viability or myogenic differentiation assays.
Other relevant examples include alginate beads functionalized with extracellular matrix proteins or peptides, where C2C12 cells are mixed into alginate, crosslinked with CaCl$_2$, and cultured for several days before cell recovery or imaging.
These examples preserve the required constraints: C2C12 cells, 3D culture, and alginate as the culture matrix.
\end{minipage}
\\

\midrule
\textbf{\textit{INF-Query-Aligner}} \\
\midrule
Explain 3D culture experiments using alginate to encapsulate C2C12 cells, focusing on cell morphology and differentiation.
\\

\midrule
\textbf{\textit{TongSearch-QR}} \\
\midrule
Find 3D culture experiments where C2C12 cells are grown in alginate.
\\

\bottomrule
\end{tabularx}
\end{table*}
\begin{table*}[!ht]
\centering
\small
\setlength{\tabcolsep}{4pt}
\renewcommand{\arraystretch}{1.1}
\caption{
Details of reranking models evaluated in \textsc{ProSearch}.
All rerankers reorder the same top-100 candidate paragraphs retrieved by Qwen3-Embedding-8B.
}
\label{tab:reranker_details}
\begin{tabular*}{\textwidth}{@{\extracolsep{\fill}}lcll@{}}
\toprule
Model & Size & Checkpoint & Ranking Format \\
\midrule
BGE-Reranker-v2-m3
& 0.6B
& BAAI/bge-reranker-v2-m3
& Pointwise \\

Rank-R1
& 7B
& ielabgroup/Rank-R1-7B-v0.1
& Setwise \\

Rank1
& 7B
& jhu-clsp/rank1-7b
& Pointwise reasoning \\

DIVER-GroupRank-7B
& 7B
& AQ-MedAI/Diver-GroupRank-7B
& Groupwise \\

ReasonRank-8B
& 8B
& lwhlwh/reasonrank-8B
& Listwise reasoning \\

Qwen3-Reranker
& 8B
& Qwen/Qwen3-Reranker-8B
& Pointwise \\
\bottomrule
\end{tabular*}
\end{table*}

\subsection{Reranking experiments}
\label{app:reranking}

We evaluate whether reranking improves protocol retrieval after first-stage dense retrieval.
For each query, we first retrieve the top-100 candidate paragraphs using Qwen3-Embedding-8B.
Each reranking model then reorders the same candidate list, and we evaluate the reranked list using nDCG@10. Table~\ref{tab:reranker_details} summarizes the reranking models and their ranking formats. Table~\ref{tab:reranker_prompts} reports the prompts used in reranking experiments.

\section{More Experimental Results and Analysis}
\label{app:experimental_detail}

\subsection{Annotation Pool Robustness}

\paragraph{Near-Miss Positive Recovery.}
To assess the effect of near-miss candidates on retrieval
evaluation, we compare the full qrels with a support-only
setting that excludes positives recovered from the near-miss
pool. Table~\ref{tab:validation_pool_effect} shows that
the average nDCG@10 changes only slightly across the
representative retrievers, with their relative ordering
largely unchanged. These results suggest that near-miss
candidates improve positive coverage without substantially
altering the retrieval conclusions.

\begin{table*}[ht]
\centering
\small
\setlength{\tabcolsep}{4pt}
\renewcommand{\arraystretch}{1.08}
\caption{
Effect of near-miss positive recovery on retrieval performance, measured by changes in nDCG@10.
Each domain column reports Full minus Support-only, where Support-only qrels include positives identified from exact-support candidates and Full qrels additionally include positives recovered from near-miss candidates.
$\Delta$ Avg. denotes the average change across the six domains.
}
\label{tab:validation_pool_effect}
\begin{tabular*}{\textwidth}{@{\extracolsep{\fill}}lccccccc@{}}
\toprule
Retriever
& Cell.
& Immuno.
& Electro.
& Organic
& Inorg.
& Polymer
& $\Delta$ Avg. \\
\midrule
BM25
& +0.002 & +0.000 & +0.001 & -0.010 & +0.005 & +0.003
& 0.000 \\

GTE-large
& +0.031 & +0.001 & +0.005 & -0.008 & +0.024 & +0.010
& +0.011 \\

ChEmbed-prog
& +0.019 & +0.004 & -0.002 & -0.015 & -0.009 & +0.002
& 0.000 \\

BGE-Reasoner-Embed
& +0.034 & +0.003 & -0.003 & -0.001 & -0.012 & +0.003
& +0.004 \\

NV-Embed-v2
& +0.053 & +0.003 & -0.005 & +0.004 & -0.009 & +0.019
& +0.011 \\

Qwen3-Embedding-8B
& +0.063 & +0.004 & -0.002 & -0.010 & -0.033 & +0.045
& +0.011 \\

\bottomrule
\end{tabular*}
\end{table*}

\paragraph{Pool-bias Probe.}
To examine whether the annotation pool misses relevant
paragraphs, we sample 300 queries (50 per domain) and
collect the top-10 results from all 19 retrievers.
We judge previously unannotated query--paragraph pairs
and recompute nDCG@10 using the expanded qrels.
As shown in Table~\ref{tab:pool_bias_probe},
only 1.86\% of newly judged candidates are relevant.
The average nDCG@10 across the 19 retrievers increases
by 0.025, while their rankings remain highly correlated
(Spearman's $\rho=0.984$).

\begin{table}[ht]
\centering
\small
\setlength{\tabcolsep}{5pt}
\renewcommand{\arraystretch}{1.08}
\caption{
Pool-bias probe on 300 queries, with 50 sampled from
each domain. Added Pos. denotes the percentage of
newly judged candidates labeled relevant.
$\Delta$ nDCG@10 is the average score change across
19 retrievers after expanding the qrels.
$\rho$ denotes the Spearman correlation between
retriever rankings before and after expansion.
}
\label{tab:pool_bias_probe}

\begin{tabular*}{\textwidth}{@{\extracolsep{\fill}}lccc@{}}
\toprule
Domain & Added Pos. (\%) & $\Delta$ nDCG@10 & $\rho$ \\
\midrule
Cell Culture        & 2.74 & +0.042 & 0.993 \\
Immunotechnology    & 1.89 & +0.029 & 0.954 \\
Electrochemistry    & 3.97 & +0.059 & 0.977 \\
Organic Synthesis   & 1.60 & +0.012 & 0.993 \\
Inorganic Materials & 0.74 & +0.009 & 0.988 \\
Polymer Materials   & 0.21 & -0.001 & 0.996 \\
\midrule
Average             & 1.86 & +0.025 & 0.984 \\
\bottomrule
\end{tabular*}
\end{table}

\subsection{Relevance Label Robustness}
\label{app:relevance_label_robustness}

\paragraph{Validator Sensitivity.}
To assess the sensitivity of relevance labels to the choice
of LLM validator, we sample 300 queries (50 per domain)
and re-judge all 16,700 query--paragraph pairs in their
original annotation pools using Qwen3-235B and
Claude Sonnet 4.6.
As shown in Table~\ref{tab:validator_sensitivity},
their agreements with GPT-5.4 are 66.5\% and 90.2\%,
respectively. Despite these differences, retriever rankings
remain highly correlated with those obtained using GPT-5.4
labels, with Spearman's $\rho=0.970$ and $0.982$.
Disagreements are more pronounced for numeric constraints
and queries containing four or more constraints.

\paragraph{Consensus Qrels.}
To reduce dependence on a single validator, we construct
consensus qrels using majority (2-of-3) and unanimous
(3-of-3) agreement among GPT-5.4, Qwen3-235B, and
Claude Sonnet 4.6.
Table~\ref{tab:consensus_qrels} shows that both settings
produce only small changes in average nDCG@10,
while retriever rankings remain stable
(Spearman's $\rho=0.985$ and $0.991$, respectively).
These results further support the robustness of the
retriever comparisons to relevance-label variation.

\paragraph{Same-family Influence.}
We further examine whether using a validator from the same
model family favors the corresponding retriever.
When switching from GPT-5.4 to Qwen3-235B labels,
Qwen3-Embedding-8B's average nDCG@10 decreases by
0.0241, compared with an average decrease of 0.0027
for the other 18 retrievers.
This comparison provides no evidence of same-family
favoritism in relevance labeling, although it does not
rule out biases introduced at other construction stages.

\begin{table}[t]
\centering
\small
\setlength{\tabcolsep}{4pt}
\renewcommand{\arraystretch}{1.08}
\caption{
Sensitivity to relevance-label validators on 300 queries
(16,700 query--paragraph pairs).
Agreement and Cohen's $\kappa$ are measured against GPT-5.4 labels.
$\rho$ denotes the Spearman correlation of the 19-retriever rankings,
and $\Delta$ Avg. denotes the change in cross-domain average nDCG@10
relative to GPT-5.4 qrels.
}
\label{tab:validator_sensitivity}

\begin{tabular*}{\linewidth}{@{\extracolsep{\fill}}lrrrr@{}}
\toprule
Validator & Agreement & $\kappa$ & $\rho$ & $\Delta$ Avg. \\
\midrule
Qwen3-235B        & 66.48\% & 0.2592 & 0.970 & $-0.0043$ \\
Claude Sonnet 4.6 & 90.20\% & 0.6353 & 0.982 & $-0.0122$ \\
\bottomrule
\end{tabular*}
\end{table}

\begin{table*}[t]
\centering
\small
\setlength{\tabcolsep}{5pt}
\renewcommand{\arraystretch}{1.08}
\caption{
Effect of consensus relevance labels on retrieval evaluation.
Majority and Unanimous qrels require agreement from at least
two and all three validators, respectively
(GPT-5.4, Qwen3-235B, and Claude Sonnet 4.6).
GPT-5.4 reports mean nDCG@10 across 19 retrievers in each domain;
$\Delta$ denotes the change from GPT-5.4 qrels, and
$\rho$ denotes the Spearman correlation of retriever rankings.
}
\label{tab:consensus_qrels}

\begin{tabular*}{\textwidth}{@{\extracolsep{\fill}}lrrrrr@{}}
\toprule
& & \multicolumn{2}{c}{Majority (2/3)}
    & \multicolumn{2}{c}{Unanimous (3/3)} \\
\cmidrule(lr){3-4}
\cmidrule(l){5-6}
Domain & GPT-5.4 & $\Delta$ & $\rho$ & $\Delta$ & $\rho$ \\
\midrule
Cell Culture        & 0.226 & $-0.008$ & 0.995 & $-0.018$ & 0.991 \\
Immunotechnology    & 0.026 & $-0.004$ & 0.951 & $0.000$  & 0.993 \\
Electrochemistry    & 0.130 & $-0.021$ & 0.995 & $-0.002$ & 0.997 \\
Organic Synthesis   & 0.171 & $+0.012$ & 0.998 & $+0.002$ & 0.997 \\
Inorganic Materials & 0.341 & $-0.004$ & 0.979 & $0.000$  & 0.982 \\
Polymer Materials   & 0.157 & $-0.001$ & 0.995 & $-0.006$ & 0.988 \\
\midrule
Average             & 0.175 & $-0.004$ & 0.985 & $-0.004$ & 0.991 \\
\bottomrule
\end{tabular*}
\end{table*}

\subsection{Method vs. Full-text Retrieval}

Table~\ref{tab:paragraph_pool_scope_all} reports the domain-wise nDCG@10 using method-only and all-body paragraph pools.
\begin{table*}[ht]
\centering
\small
\setlength{\tabcolsep}{4pt}
\renewcommand{\arraystretch}{1.08}
\caption{
Comparison of retrieval performance over method paragraphs and all body paragraphs across six domains.
Each domain column reports the change in nDCG@10 (All-body minus Method-only), and $\Delta$ Avg. denotes the average change across the six domains.
}
\label{tab:paragraph_pool_scope_all}
\begin{tabular*}{\textwidth}{@{\extracolsep{\fill}}lccccccc@{}}
\toprule
Retriever
& Cell.
& Immuno.
& Electro.
& Organic
& Inorg.
& Polymer
& $\Delta$ Avg. \\
\midrule
BM25
& -0.048 & -0.005 & -0.029 & -0.071 & -0.078 & -0.108
& -0.057 \\

BGE-Reasoner-Embed
& -0.091 & -0.008 & -0.045 & -0.107 & -0.067 & -0.096
& -0.069 \\

NV-Embed-v2
& -0.095 & -0.010 & -0.056 & -0.109 & -0.070 & -0.140
& -0.080 \\

Qwen3-Embedding-8B
& -0.080 & -0.026 & -0.047 & -0.102 & -0.121 & -0.128
& -0.084 \\

\bottomrule
\end{tabular*}
\end{table*}

\subsection{Query Characteristics}
\label{query_characteristics}

\begin{figure}[t]
    \centering
    \includegraphics[width=\linewidth]{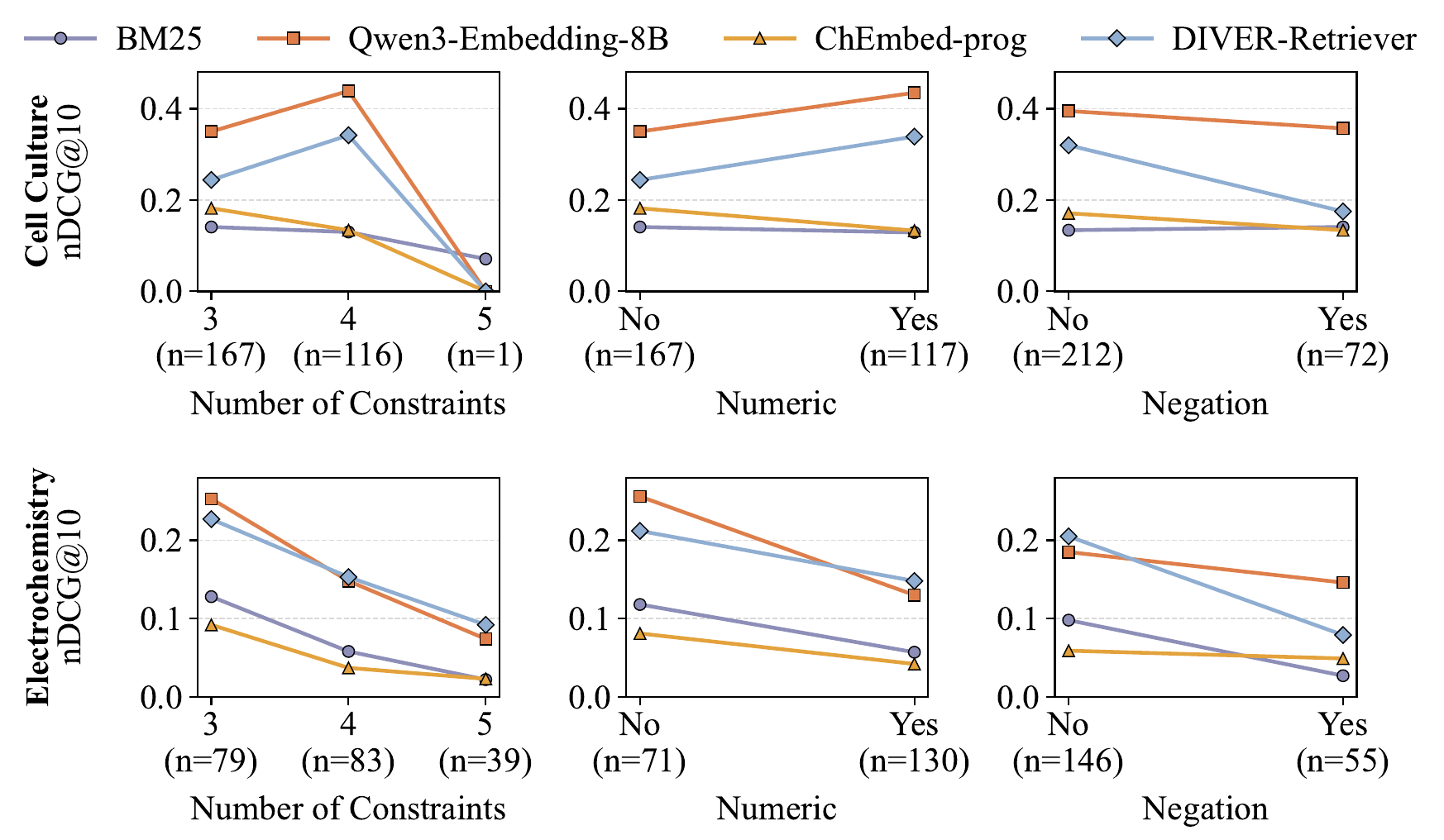}
    \caption{
    Retrieval performance by query characteristics in Cell Culture and Electrochemistry. Queries are grouped by constraint count, numeric constraints, and negation constraints.
    }
    \label{fig:diagnostic_retrieval_analysis}
\end{figure}

We analyze retrieval performance by constraint count, numeric
and negation constraints, and constraint roles.
Figure~\ref{fig:diagnostic_retrieval_analysis} compares
representative retrievers in Cell Culture and Electrochemistry.

We further examine constraint roles, focusing on
\textit{Condition} and \textit{Method}, since \textit{Goal}
appears in every query and \textit{Input} in 97.0\%.
As shown in Table~\ref{tab:role_level_analysis},
queries containing \textit{Condition} constraints achieve
lower average nDCG@10 for both BM25 and Qwen3-Embedding-8B,
whereas \textit{Method} constraints show no comparable decrease.

\begin{table}[t]
\centering
\small
\setlength{\tabcolsep}{5pt}
\renewcommand{\arraystretch}{1.08}
\caption{
Retrieval performance grouped by constraint role, measured by average nDCG@10 across the six domains. For each role (Condition and Method), queries are divided into two groups according to whether they contain constraints assigned to that role (Present) or not (Absent). Qwen3 refers to Qwen3-Embedding-8B.
}
\label{tab:role_level_analysis}

\begin{tabular*}{\textwidth}{@{\extracolsep{\fill}}lcccc@{}}
\toprule
& \multicolumn{2}{c}{Condition}
& \multicolumn{2}{c}{Method} \\
\cmidrule(lr){2-3}
\cmidrule(l){4-5}
Retriever & Absent & Present & Absent & Present \\
\midrule
BM25  & 0.181 & 0.136 & 0.132 & 0.159 \\
Qwen3 & 0.356 & 0.285 & 0.288 & 0.293 \\
\bottomrule
\end{tabular*}
\end{table}

\subsection{Manual error analysis examples}
\label{app:error_examples}

Table~\ref{tab:error_analysis_examples} provides representative examples from the manual error analysis.

\section{Prompts and Dataset Examples}
\label{app:prompts_examples}

Tables~\ref{tab:example_cell_culture}--\ref{tab:example_polymer_materials}
provide representative query examples from the six domains in ProSearch.
Each example includes the natural-language query, its structured constraints, a positive paragraph, and a negative paragraph.
These examples illustrate how relevance in ProSearch depends on satisfying all query constraints within the same experimental setup.

Tables~\ref{tab:prompt_record_extraction} and~\ref{tab:prompt_value_normalization} show the prompts for extracting structured protocol records from method paragraphs and normalizing the extracted field values, respectively.
Tables~\ref{prompt_fixed_constraint_query_generation}, 
\ref{prompt:exploration_query_generation}, and~\ref{prompt:query_quality_filtering} present the prompts for query generation and query quality filtering.
Tables~\ref{prompt:fixed_constraint_relevance_judging} and~\ref{prompt:exploration_relevance_judging} show the prompts used for judging whether a paragraph satisfies the query constraints.

\section{Annotation Details}

We provide the expert annotation instructions used for query quality evaluation and relevance label verification in Tables~\ref{tab:expert_query_annotation_instruction} and~\ref{tab:expert_label_annotation_instruction}.

\begin{table*}[ht]
\centering
\footnotesize
\setlength{\tabcolsep}{3pt}
\renewcommand{\arraystretch}{1.12}

\caption{
Representative examples from the manual error analysis of Qwen3-Embedding-8B. Each example includes the query, a retrieved paragraph excerpt, the assigned error category, and the reason why the paragraph fails to satisfy the query constraints. Paragraphs are shortened for readability.
}
\label{tab:error_analysis_examples}
\begin{tabularx}{\textwidth}{@{}p{0.19\textwidth}X p{0.12\textwidth} p{0.22\textwidth}@{}}
\toprule
Query & Retrieved paragraph excerpt & Error category & Reason \\
\midrule

Find 3D culture conditions for HCT116 cells in DMEM.
&
The HCT116 colon cancer cell line was collected for cell culture. RPMI1640 supplemented with 10\% fetal bovine serum was prepared at 37$^\circ$C. Tumor spheres of HCT116 cells were formed in suspension. Sphere Formation Media was composed of RPMI1640, epidermal growth factor, fibroblast growth factor-basic, B27, and bovine serum albumin. The cells were diluted to 2 $\times$ 10$^4$/ml and cultured in 24-well plates.
&
\makecell{Categorical}
&
The paragraph describes HCT116 sphere culture, but the medium is RPMI/SFM rather than DMEM.
\\

\midrule
Find 3D culture conditions for HCT116 cells in DMEM.
&
HCT 116 cell lines were maintained in DMEM supplemented with 10\% FBS, L-glutamine, penicillin, and streptomycin. HCT 116 cells were detached by trypsin-EDTA. Cells were cultivated at 37$^\circ$C, 5\% CO$_2$, and 95\% humidity.
&
\makecell{ Categorical}
&
The paragraph gives DMEM culture conditions for HCT116 cells, but does not describe a 3D culture setup.
\\

\midrule
What basal medium other than RPMI is used for K-562 cells cultured in suspension?
&
The continuous cell line K562 was established by Lozzio and Lozzio. The suspension cell lines were cultivated using Roswell Park Memorial Institute (RPMI) 1640. The medium was supplemented with 10\% fetal bovine serum and 1\% Penicillin-Streptomycin. The cells were incubated at 37$^\circ$C and 5\% CO$_2$.
&
\makecell{Negation}
&
The query asks for a basal medium other than RPMI, but the paragraph explicitly uses RPMI 1640.
\\

\midrule
What solvent other than DCM is used for Heck cross-coupling with Pd@ASP-EDTA-CS?
&
Reaction conditions: aryl halide, alkene, K$_2$CO$_3$, Pd@ASP-EDTA-CS, and solvent unless otherwise stated. Isolated yield.
&
\makecell{Topic-only}
&
The paragraph is about the relevant Heck reaction system, but it does not name a concrete solvent value.
\\

\bottomrule
\end{tabularx}
\end{table*}
\begin{table*}[t]
\centering
\footnotesize
\setlength{\tabcolsep}{0pt}
\renewcommand{\arraystretch}{1.12}
\caption{
An illustrative example from the cell culture domain, showing a natural-language query, its corresponding structured constraints, and paired positive and negative method paragraphs.
}
\label{tab:example_cell_culture}
\begin{tabularx}{\textwidth}{@{}X@{}}
\toprule

\textbf{\textit{Query}} \\
\midrule
Find 3D culture experiments in which C2C12 cells are grown in alginate.
\\

\midrule
\textbf{\textit{Structured Constraints}} \\
\midrule
\begin{tabular}{@{}lll@{}}
\texttt{culture\_mode} & $\;=\;$ & \texttt{3D} \\
\texttt{culture\_substrate\_or\_coating} & $\;=\;$ & \texttt{alginate} \\
\texttt{cultured\_cells.cell\_type} & $\;=\;$ & \texttt{C2C12} \\
\end{tabular}
\\

\midrule
\textbf{\textit{Example Positive Paragraph}} \\
\midrule
The voltage of the electrostatic droplet generator was optimized and two commercial cell attachment factors (gelatin and ECL) were tested to establish robust production of alginate microcapsules. A voltage of 5--5.25 kV for droplet generation was applied, resulting in spherical microcapsules and a homogenous size distribution ($\pm$2--3\%) (Supplementary Table S1). Furthermore, we explored the incorporation of gelatin and ECL into the alginate as a cell adhesion matrix in both serum (DMEM+FBS) and serum-free media (AIM-V), since alginate lacks peptide sequence that supports cell attachment. Overall, encapsulated C2C12 cells in alginate microcapsule cultured in serum-free media showed a decreasing trend in viability, but no significant difference, compared to that cultured in serum-supplemented media, demonstrating the potential of serum-free media for the alginate microcapsule cultivation. A significant difference in cell viability between the addition of two extracellular matrices (ECL vs. Gelatin) into alginate was observed. Alginate beads with ECL showed significantly higher cell viability than alginate alone and alginate with gelatin in both serum-supplemented and serum-free media cultivation (Supplementary Figure S1). Finally, the addition of ECL was chosen to enhance cell survival in alginate microcapsule culture systems.
\\

\midrule
\textbf{\textit{Example Negative Paragraph}} \\
\midrule
Response surface method (RSM). Based on the screening of the O$_2$ strategies, RSM was used to optimize the BAP configuration concerning the O$_2$ balance in the alginate sheet device. The objective was to maximize the density of viable MPIs in the BAP by tuning the HEMOXCell concentration and the islet seeding density in the hypoxic environment in the presence of the silicone-CaO$_2$ disk. A central composite design ($2^2$ factorial design with 4-star points and four replicates of the central point) was used to fit a second-order polynomial model (Fig. 2C). The model validation was performed by analyzing the lack-of-fit test results, the determination coefficient R$^2$ value, and the diagnostic plots. The two factors studied were HEMOXCell concentration and islet seeding density, and their ranges were determined according to the literature and the determination of silicone-CaO$_2$ OTR and pseudo-islet OCR. The response variables were intracellular ATP content (RLU) and ATP/LDH viability ratio (RLU/AU) per device. These response variables were assessed on pseudo-islets encapsulated in alginate sheets cultured for 24 h under the different conditions defined by the experimental plan (Table 5). The optimum values were determined by solving the regression equations and analyzing the response surface plots. A multi-response optimization was performed to achieve the best compromise to maximize viability and function of the encapsulated pseudo-islets.
\\

\bottomrule
\end{tabularx}
\end{table*}
\begin{table*}[t]
\centering
\footnotesize
\setlength{\tabcolsep}{0pt}
\renewcommand{\arraystretch}{1.12}
\caption{
An illustrative example from the immunotechnology domain, showing a natural-language query, its corresponding structured constraints, and paired positive and negative method paragraphs.
}
\label{tab:example_immunotechnology}
\begin{tabularx}{\textwidth}{@{}X@{}}
\toprule

\textbf{\textit{Query}} \\
\midrule
What loading amount is used for Western blot detection of BECN1 in tissue protein extract?
\\

\midrule
\textbf{\textit{Structured Constraints}} \\
\midrule
\begin{tabular}{@{}lll@{}}
\texttt{protein\_targets.target} & $\;=\;$ & \texttt{BECN1} \\
\texttt{sample\_type} & $\;=\;$ & \texttt{tissue protein extract} \\
\texttt{loading\_amount} & $\;\neq\;$ & \texttt{null} \\
\end{tabular}
\\

\midrule
\textbf{\textit{Example Positive Paragraph}} \\
\midrule
Tibial tissues were lysed with radioimmunoprecipitation assay buffer (Cell Signaling Technology, Danvers, MA) containing protease inhibitor cocktail. The concentration of proteins (10 $\mu$g per lane) was determined using a Pierce\texttrademark\ BCA Protein Assay Kit. Proteins were separated by sodium dodecyl sulfate-polyacrylamide gel electrophoresis and were then transferred to polyvinylidene difluoride (PVDF) membranes at 15 V for 1 h at low temperature. Membranes were blocked with 5\% skim milk for 2 h. Proteins were detected using the following primary antibodies at a 1:1000 dilution overnight at 4~$^\circ$C: anti-Beclin-1 (ab210498; Abcam, UK), anti-light chain 3B (LC3B; ab48394; Abcam), anti-GNB1 (ab137635; Abcam), anti-autophagy-related 5 (ATG5; ab108327; Abcam), anti-p62 (ab56416; Abcam), anti-Bcl-2 (ab194583; Abcam), anti-Bax (ab32503; Abcam), anti-cleaved caspase 3 (PA5-114687, Thermo Fisher Scientific) and anti-GAPDH (ab9485; Abcam) antibodies. Then, membranes were incubated with horseradish peroxidase-conjugated antirabbit IgG (Zsbio) for 2 h at room temperature. Finally, immunoreactions were evaluated with an enhanced chemiluminescence detection system. The protein levels were normalized to those of the corresponding GAPDH loading control. An Alpha Innotech Gel Imager (Alpha Innotech Corporation, San Leandro, CA) was used to visualize the specific bands, and the optical density of each band was measured using Image Pro Plus software (version X; Media Cybernetics, Silver Springs, MD).
\\

\midrule
\textbf{\textit{Example Negative Paragraph}} \\
\midrule
The total protein yield in the TMJ periarticular tissues was measured through the BCA protein assay kit (Thermo Scientific, Rockford, IL, USA). Protein samples (80 $\mu$g) of periarticular tissues were separated on SDS/PAGE gel and transferred for nitrocellulose membranes. A molecular mass standard (Bio-Rad, Hercules, CA, USA) was run in parallel to estimate molecular mass. The blockade of the membranes was performed in TBST (Tris-Buffered Saline, 0.1\% Tween 20) containing 5\% of nonfat milk at 4~$^\circ$C overnight. The membranes were washed in TBS and then incubated at room temperature with primary antibodies: iNOS (M1 macrophage) antibody (overnight, 1:500, \#18985-1-AP, Proteintech, Rosemont, IL, USA), MRC1 (M2 macrophage) antibody (overnight, 1:500, \#18704-1-A, Proteintech, Rosemont, IL, USA) or GAPDH antibody (2 h, 1:1000, \#5174, Cell Signaling Technology, Danvers, MA, USA), used as a housekeeping protein. The membranes were then rewashed and incubated (2 h) with the appropriate secondary antibody conjugated with peroxidase (1:10000, \#A6667, Sigma-Aldrich, St. Louis, MO, USA). The target protein was visualized in the membrane using a chemiluminescence-based ECL system (Amersham Biosciences, Piscataway, NJ, USA) and the digital image was obtained by CCD camera imaging for chemiluminescence (ImageQuant LAS 4000 mini, GE Healthcare Life Sciences, Pittsburgh, PA, USA). The program Image J (National Institutes of Health, Bethesda, USA) was utilized to measure the optical density of the bands.
\\

\bottomrule
\end{tabularx}
\end{table*}
\begin{table*}[t]
\centering
\footnotesize
\setlength{\tabcolsep}{0pt}
\renewcommand{\arraystretch}{1.12}
\caption{
An illustrative example from the electrochemistry domain, showing a natural-language query, its corresponding structured constraints, and paired positive and negative method paragraphs.
}
\label{tab:example_electrochemistry}
\begin{tabularx}{\textwidth}{@{}X@{}}
\toprule

\textbf{\textit{Query}} \\
\midrule
Find cyclic voltammetry experiments targeting DCF in phosphoric acid electrolyte.
\\

\midrule
\textbf{\textit{Structured Constraints}} \\
\midrule
\begin{tabular}{@{}lll@{}}
\texttt{electrolytes.chemical} & $\;=\;$ & \texttt{phosphoric acid} \\
\texttt{measurement\_method} & $\;=\;$ & \texttt{CV} \\
\texttt{target\_species} & $\;=\;$ & \texttt{DCF} \\
\end{tabular}
\\

\midrule
\textbf{\textit{Example Positive Paragraph}} \\
\midrule
The Cyclic Voltammetry (CV) measurements were carried out by GCE and aGCE and by SPCE and aSPCE to study the DCF electrochemical behavior. With the aGCEs, the following measurement parameters were used: time for equilibration: 15 s; potential step: 5.0 mV; potential range: 0.0--0.9 V; scan rate: 0.02--0.175 V s$^{-1}$. The CV signals were recorded on solution $7\times10^{-8}$ mol L$^{-1}$ DCF at different pH and medium: pH 2.0, with $1\times10^{-2}$ mol L$^{-1}$ hydrochloric acid; pH 2.0, with $2.5\times10^{-2}$ mol L$^{-1}$ phosphoric acid; pH 4.7, with $2.5\times10^{-2}$ mol L$^{-1}$ acetic buffer; pH 7.0, with $2.5\times10^{-2}$ mol L$^{-1}$ phosphate buffer.
\\

\midrule
\textbf{\textit{Example Negative Paragraph}} \\
\midrule
All chemicals used in this study were of analytical grade, supplied from Sigma Aldrich. All reagents were used as purchased without any further treatment. For the preparation of the solutions ultrapure MilliQ water was used. Doxorubicin hydrochloride was prepared in the concentration of 1 mM for cyclic voltammetry (CV) measurements and from this solution standard solutions for quantification were prepared. Britton--Robinson buffer solution (BRBS) was used as supporting electrolyte. BRBS was prepared with 0.04 M sulfuric acid, 0.04 M boric acid and 0.04 M phosphoric acid. 0.2 M sodium hydroxide was used to adjust higher pH values of BRBS.
\\

\bottomrule
\end{tabularx}
\end{table*}
\begin{table*}[t]
\centering
\footnotesize
\setlength{\tabcolsep}{0pt}
\renewcommand{\arraystretch}{1.12}
\caption{
An illustrative example from the organic synthesis domain, showing a natural-language query, its corresponding structured constraints, and paired positive and negative method paragraphs.
}
\label{tab:example_organic_synthesis}
\begin{tabularx}{\textwidth}{@{}X@{}}
\toprule

\textbf{\textit{Query}} \\
\midrule
What additive is used in Suzuki-Miyaura coupling with Pd(OAc)$_2$?
\\

\midrule
\textbf{\textit{Structured Constraints}} \\
\midrule
\begin{tabular}{@{}lll@{}}
\texttt{catalysts.name} & $\;=\;$ & \texttt{Pd(OAc)2} \\
\texttt{transformation\_type} & $\;=\;$ & \texttt{Suzuki-Miyaura coupling} \\
\texttt{additives} & $\;\neq\;$ & \texttt{null} \\
\end{tabular}
\\

\midrule
\textbf{\textit{Example Positive Paragraph}} \\
\midrule
General method for the synthesis of 1,1-diarylvinyl sulfones CA4 analogues (4a--u) [48][49][50]. To a stirred mixture of (E)-5-(1-iodo-2-tosylvinyl)-1,2,3-trimethoxybenzene 4A (100 mg, 0.211 mmol, 1 eq), different aryl boronic acids (30.64 mg, 0.253 mmol, 1.2 eq) and Na$_2$CO$_3$ (44.73 mg, 0.422 mmol, 2 eq) in water and dimethylformalmide (1:2), with palladium acetate (Pd(OAc)$_2$) catalyst (0.0006 mmol, 0.005 eq) was added and reaction was stirred at room temperature under nitrogen for 3 h. The residue was purified by chromatography (EtOAc/hexanes) to afford the corresponding coupled product 4A-V (Fig. 4d). Antiproliferative assay. The cytotoxicity of synthesized compounds 4A-V were determined in triplicates by 3-[4,5-dimethylthiazol-2-yl]-2,5-diphenyltetrazolium bromide (MTT) assay [51,52]. The pale yellow coloured tetrazolium salt (MTT) reduced to a dark blue water-insoluble formazan by metabolically active cells and the product was measured quantitatively after solubility in DMSO. The absorbance of the soluble formazan is directly proportional to the number of viable cells. A panel of four cancer cell lines were used for testing the in-vitro cytotoxicity. These cell lines were grown in Dulbecco's modified Eagle's medium (DMEM) containing non-essential amino acids and 10\% FBS. All the cells were maintained under humidified conditions of 5\% CO$_2$ atmosphere at 37~$^\circ$C in a CO$_2$ incubator (Model Galaxy 170S, Eppendorf, USA). The 96-well micro titre plates were incubated for 24 h prior to addition of the experimental compounds. Cells were treated with vehicle alone (DMSO) or compounds (drugs (1 $\mu$g/mL) were dissolved in 100 $\mu$L DMSO previously) at different concentrations (0.1, 1, 10 and 25 $\mu$M) of test compounds for 48 h. The assay was completed with the addition of MTT (5\%, 10 $\mu$L) and incubated for 60 min at 37~$^\circ$C. The supernatant was aspirated and plates were air dried and the MTT-formazan crystals were dissolved in 100 $\mu$L of DMSO.
\\

\midrule
\textbf{\textit{Example Negative Paragraph}} \\
\midrule
To an oven-dried 35 mL Schlenk tube with previously placed magnetic stir-bar were added aryl iodide 2 (0.5 mmol, 2.5 equiv.), aryl boronic acid (0.1 mmol, 0.5 equiv.), Pd(OAc)$_2$ (4.5 mg, 0.02 mmol, 10 mol\%), Fmoc-Phg-OH (22.4 mg, 0.06 mmol, 30 mol\%), Ag$_2$CO$_3$ (110 mg, 0.4 mmol, 2 equiv.), followed by addition of HFIP (4 mL) and alkene substrate 1 (0.2 mmol). The tube was sealed with a screw cap and the reaction mixture was stirred vigorously on a hotplate at 45~$^\circ$C for 24 h. After completion of the reaction, the resultant solution was filtered through a short pad of 1:1 mixture of Celite and silica gel, and the column was washed with ethyl acetate (15 mL). The combined organic solutions were concentrated under reduced pressure, and the residue was purified by flash column chromatography on silica gel to afford the desired product 5.
\\

\bottomrule
\end{tabularx}
\end{table*}
\begin{table*}[t]
\centering
\footnotesize
\setlength{\tabcolsep}{0pt}
\renewcommand{\arraystretch}{1.12}
\caption{
An illustrative example from the inorganic materials domain, showing a natural-language query, its corresponding structured constraints, and paired positive and negative method paragraphs.
}
\label{tab:example_inorganic_materials}
\begin{tabularx}{\textwidth}{@{}X@{}}
\toprule

\textbf{\textit{Query}} \\
\midrule
What sintering time is used for MgO-doped mullite?
\\

\midrule
\textbf{\textit{Structured Constraints}} \\
\midrule
\begin{tabular}{@{}lll@{}}
\texttt{dopant\_or\_additive.identity} & $\;=\;$ & \texttt{MgO} \\
\texttt{target\_material} & $\;=\;$ & \texttt{mullite} \\
\texttt{sintering.time} & $\;\neq\;$ & \texttt{null} \\
\end{tabular}
\\

\midrule
\textbf{\textit{Example Positive Paragraph}} \\
\midrule
Different amounts of MgO were added to the mullite stoichiometric composition 
(71.8 wt\% of Al$_2$O$_3$ and 28.2 wt\% of SiO$_2$), as follows: 
0, 0.8, 1.6, 2.4, 3.1, and 4.0 wt\%. These compositions were identified as M0, M1, M2, M3, M4, and M5, respectively. The compositions were mixed in a ball mill for 6 h. Next, they were dried with hot air flow and deagglomerated in an 80 mesh nylon sieve. To evaluate the densification of the compositions with different MgO amounts, dilatometry tests were carried out. For the tests, a sample of each composition was formed using uniaxial pressing, in a cylindrical mold. From the densification behavior, the determined sintering temperatures were used. The dilatometry was performed up to 1400~$^\circ$C at a 10~$^\circ$C/min heating rate, in an air atmosphere, with a dilatometer (Setsys Evolution, Setaram). To study the material sinterability, the samples were formed similarly to those used in the dilatometry. Then, the green samples were sintered at a 10~$^\circ$C/min constant heating rate up to the sintering temperature (1350, 1450, and 1550~$^\circ$C), with a 4 h holding time.
\\

\midrule
\textbf{\textit{Example Negative Paragraph}} \\
\midrule
The microstructures of alumina samples sintered at various sintering temperatures are depicted in Fig. 3. The effect of the sintering temperature, amount of waste alumina, and grain size on the properties of alumina samples is demonstrated in Table 3. The value of dwell time was fixed at 2 h because of the negligible impact on density, hardness, and indentation fracture toughness in the investigated range. At the lowest investigated sintering temperature of 1400~$^\circ$C, the consolidation process was incomplete; therefore, the so-called necking between alumina particles can be seen and pronounced grain boundaries were absent as well, as shown in Fig. 3A and 3B. This has been reflected in very low hardness and indentation fracture toughness values (Table 3). The increase in the sintering temperature to 1600~$^\circ$C increased the consolidation rate, resulting in distinct and visible grain boundaries. The increase in obtained mechanical properties is also noticed (Table 3). The increasing content of the waste alumina in studied samples resulted in finer microstructure due to the higher amount of MgO (Table 1). Magnesia doping causes a reduction in grain boundary mobility of alumina, resulting in the reduction in alumina grain size at higher sintering temperatures (Fig. 3C and 3D) [17,18]. The addition of 20 dwb.\% of waste alumina powder did not negatively affect the obtained density (Table 2) and microstructure of sintered samples (Fig. 3).
\\

\bottomrule
\end{tabularx}
\end{table*}
\begin{table*}[t]
\centering
\footnotesize
\setlength{\tabcolsep}{0pt}
\renewcommand{\arraystretch}{1.12}
\caption{
An illustrative example from the polymer materials domain, showing a natural-language query, its corresponding structured constraints, and paired positive and negative method paragraphs.
}
\label{tab:example_polymer_materials}
\begin{tabularx}{\textwidth}{@{}X@{}}
\toprule

\textbf{\textit{Query}} \\
\midrule
What PEGDA loading is used with alginate as the polymer system?
\\

\midrule
\textbf{\textit{Structured Constraints}} \\
\midrule
\begin{tabular}{@{}lll@{}}
\texttt{crosslinking\_agent.name} & $\;=\;$ & \texttt{PEGDA} \\
\texttt{polymer\_system} & $\;=\;$ & \texttt{alginate} \\
\texttt{crosslinking\_agent.loading} & $\;\neq\;$ & \texttt{null} \\
\end{tabular}
\\

\midrule
\textbf{\textit{Example Positive Paragraph}} \\
\midrule
Effects of precursor compositions: To examine the effect of glycerol ratios, mixed solvents were prepared with glycerol mass ratios ($g$) of 25\%, 40\%, 50\%, and 60\%, dissolving alginate at a concentration of 1.128\% (w/w) in each. Acrylamide monomers (19.33\% w/w) and PEGDA crosslinkers (0.67\% w/w, $m = 3.497 \times 10^{-3}$) were added, and the mixtures were stirred overnight to obtain a clear solution. The solutions were divided into degassed and non-degassed groups. For modulus measurements, 3 g of each solution was sonicated for 2, 3, 4, 6, and 8 min at 53 W cm$^{-2}$. For fracture energy measurements, 3 g of solution was sonicated for 6 min at the same intensity, followed by post-crosslinking as described above.
\\

\midrule
\textbf{\textit{Example Negative Paragraph}} \\
\midrule
Materials. Clevios PH 1000 PEDOT:PSS (1.3 wt\%) aqueous solution was purchased from Heraeus. Ceramic 3Dresyn CDP-WS Water-Soluble and Fine Turner FT2 ultrafast photoinitiator were supplied by 3Dresyns. Poly(ethylene glycol) diacrylate (PEGDA, $M_n = 250$, 575, and 700 Da) and ethylene glycol (EG) were provided by Sigma-Aldrich. Photopolymerizable and conductive hydrogels. Photopolymerizable conductive ink. The photocurable conductive ink was prepared by mixing PEDOT:PSS aqueous solution (50 wt\%) with 3Dresyn CDP-WS (34 wt\%), Fine Turner FT2 (4 wt\%), EG (4 wt\%), and PEGDA (8 wt\%) of different $M_n$ (250, 575, and 700 Da). Samples were named as PEDOT $x$\_PEGDA $n$ where $x$ refers to PEDOT percentage and $n$ to the $M_n$ of PEGDA. The mixture was protected with aluminum paper from light and stirred for 24 h. Photopolymerization reaction kinetics. The photopolymerization reaction conditions were studied by Fourier transform infrared spectroscopy (FTIR) and photorheology.
\\

\bottomrule
\end{tabularx}
\end{table*}

\begin{table*}[t]
    \centering
    \small
    \caption{
    Prompt for protocol record extraction.
    }
    \label{tab:prompt_record_extraction}
    \begin{adjustbox}{max width=1\linewidth}
    \begin{tabular}{p{\linewidth}}
    \toprule
    \begin{minipage}{0.98\linewidth}
\begin{lstlisting}[style=promptstyle]
You are a professional information extractor specializing in {domain_name}.
|split|
Your task is to determine whether the paragraph contains at least one valid experimental record for this domain and, if it does, to extract one or more setup-level structured records.

A paragraph is valid only if it provides at least one piece of experimental information that can be mapped to the defined schema fields. This information may be stated explicitly or be recoverable through short, evidence-grounded domain reasoning. A paragraph is invalid if it mainly focuses on {irrelevant_def} and does not provide any extractable value for the defined fields.

If the paragraph is invalid, return:
{{"is_valid": false, "invalid_reason": one label from {invalid_labels}, "setups": []}}

If the paragraph is valid, return:
{{"is_valid": true, "invalid_reason": null, "setups": [...]}}

Use SETUP as the extraction unit. Create a new setup only when the paragraph explicitly describes a different system, composition, condition set, treatment, reaction setting, measurement configuration, material-role assignment, or other clearly distinct experimental configuration. Do not split one setup into multiple records just because the paragraph repeats it, comments on it, or reports outcomes for it.

Extract only the fields defined in the allowed field structure below. Do not extract other experimental details. If a mention does not clearly belong to one of these fields, omit it rather than forcing it into the nearest field. The goal is high-precision extraction of a small set of structured field values, not broad extraction of all information in the paragraph. When a field is decomposed into subfields, assign each piece of information only to its own subfield and do not repeat the same content across sibling subfields. In general, the "value" should contain only the core content for that field, while "evidence" should be an exact supporting span copied from the paragraph.

Use evidence-grounded scientific reasoning when necessary to map the paragraph to the defined fields. This includes standard domain interpretations, conventional experimental terminology, and short inference steps that are strongly supported by the paragraph. However, do not invent missing information, do not use multi-step speculation, and do not replace weak hints with confident conclusions. When the inference is weak or uncertain, omit the field.

You may lightly clean the "value" when the change is obvious and low-risk, such as normalizing spaces, symbols, hyphenation, or unit formatting. However, do not change the meaning of the extracted value.

Only extract information that is concrete enough to be useful for the defined fields. If a mention is vague, generic, weakly specified, not informative as a structured field value, or not clearly matched to one of the defined fields, omit it.

The schema shown below is only a reference to the available fields and their nesting structure. It is not a template that must be filled. In the actual output, include only fields that have extracted values supported by the paragraph. If a field or subfield has no value, omit it entirely. Never output null for missing fields. Never output empty objects, empty lists, or placeholder list items.

Always include "setup_id" for each extracted setup. Use setup IDs as "setup_1", "setup_2", and so on, following order of appearance.

Return JSON only, with no extra text.

Allowed field structure: {output_schema}

Field guide: {field_glossary}

Paragraph: {paragraph}
\end{lstlisting}
    \end{minipage}
    \\
    \bottomrule
    
    \end{tabular}
    \end{adjustbox}
    
\end{table*}
\begin{table*}[t]
    \centering
    \small
    \caption{
    Prompt for field-level value normalization.
    }
    \label{tab:prompt_value_normalization}
    \begin{adjustbox}{max width=1\linewidth}
    \begin{tabular}{p{\linewidth}}
    \toprule
    \begin{minipage}{0.98\linewidth}
\begin{lstlisting}[style=promptstyle]
You are a careful scientific data normalizer.
|split|
I will give you the domain, the field name, a field-specific requirement, and a batch of items. Each item contains an item_id, a current value list, and its evidence. Your task is to normalize the current values for this field so that obvious surface variation is reduced, while distinctions that may still matter are preserved.

Use the current value list as the anchor for each item. Read the evidence only to understand, clarify, or clean those same values in context. Do not use the evidence to introduce a new value that is not already present in the current value list, and do not replace the current item with another mention that happens to appear nearby in the sentence. If the evidence contains multiple related mentions, keep only what clearly corresponds to the current item itself.

Follow the field requirement closely. Keep only content that genuinely belongs to the current field, and remove surrounding explanation, supplier information, concentrations, procedural wording, result-oriented language, and other details that do not belong in the normalized value, unless the field requirement makes that detail necessary for preserving a meaningful distinction. When a current value already expresses a clean and field-appropriate concept, keep it with only minimal normalization. When different expressions clearly refer to the same concept, use one consistent canonical form. At the same time, do not merge values that may still differ in a way that matters.

If a current value explicitly contains more than one value for the same field, you may split it into separate normalized values. But do not create extra values just because the evidence mentions other related items. If a value is clearly not appropriate for the current field, return an empty list for that item. If it may belong to the field but cannot be normalized confidently, keep a minimally cleaned version rather than making a strong inference.

Return a single JSON object only. Each key must be an item_id from the input batch, and each value must be a JSON array of canonical strings. Use an empty array when nothing should be kept. Remove duplicates within each array and sort the final values alphabetically.

Domain: {domain}
Field name: {slot_name}
Field requirement: {slot_requirement}
Batch items:
{batch_items}

Return the JSON object only.
\end{lstlisting}
    \end{minipage}
    \\
    \bottomrule
    \end{tabular}
    \end{adjustbox}
    
\end{table*}
\begin{table*}[p]
    \centering
    \small
    \caption{
    Prompt for fixed-constraint query generation.
    }
    \label{prompt_fixed_constraint_query_generation}
    \begin{adjustbox}{max width=1\linewidth}
    \begin{tabular}{p{\linewidth}}
    \toprule
    \begin{minipage}{0.98\linewidth}
\begin{lstlisting}[style=promptstyle]
You are given a structured experimental record from {domain}.

Your task is to decide whether this record should be kept as a final setup-level retrieval query candidate. If it should be kept, rewrite it as one concise retrieval request.

The request should target matching experimental evidence at the setup level, such as a method paragraph, experimental configuration, condition description, or protocol detail. Do not phrase it as searching for papers, articles, studies, documents, or literature.

Follow these steps:

1. Check whether the record defines a coherent and self-contained experimental target. The fields must work together as one meaningful experimental target. Do not judge fields one by one in isolation.

2. Skip the record if it does not form a suitable retrieval target.
   Common reasons include:
   - a required parent or anchor field is missing;
   - a field value appears unstable, non-standard, or does not fit the field meaning;
   - the field combination is methodologically awkward or internally inconsistent;
   - a numeric value, range, ratio, or loading cannot be interpreted without missing context;
   - the target is too broad, trivial, redundant, or dominated by minor protocol details;
   - the record would require adding unstated assumptions to make the query meaningful.

3. If the record is usable, write one natural retrieval request.
   The request_query must be a complete natural-language request starting with "Find".
   Prefer natural experimental wording such as:
   - "Find experiments ..."
   - "Find electrochemical experiments ..."
   - "Find immunoblotting experiments ..."
   - "Find culture conditions ..."
   - "Find synthesis protocols ..."
   - "Find fabrication protocols ..."
   - "Find polymerization experiments ..."

   Do not write a title-like noun phrase, such as "Electrochemical setups targeting O2 in 50 mM Na2SO4 electrolyte." Avoid using "setup" or "setups" in the final query unless it is clearly the most natural wording for the domain.

   Preserve all fixed constraints exactly. You may reorder information and use exact equivalents, such as standard abbreviations, standard names, formula variants, exact unit conversions, and light syntactic changes. Do not drop constraints. Do not add missing conditions. Do not broaden or narrow the target. Do not change the meaning of numbers, ranges, or negations. If an exact reformulation is uncertain, keep the original wording.

Return only one JSON object.

If the record should be kept:
{{"request_query": "<one concise retrieval request starting with Find>", "skip": false, "skip_note": null}}

If the record should be skipped:
{{"request_query": null, "skip": true, "skip_note": "<one brief sentence explaining why the record is not suitable>"}}

Do not output any explanation, note, or markdown fence.

Field definitions: {field_definitions_pretty}

Examples: {Examples}

Structured record: {record_pretty_json}
\end{lstlisting}
    \end{minipage}
    \\
    \bottomrule
    \end{tabular}
    \end{adjustbox}
    
\end{table*}
\begin{table*}[p]
    \centering
    \small
    \caption{
    Prompt for exploration query generation.
    }
    \label{prompt:exploration_query_generation}
    \begin{adjustbox}{max width=1\linewidth}
    \begin{tabular}{p{\linewidth}}
    \toprule
    \begin{minipage}{0.98\linewidth}
\begin{lstlisting}[style=promptstyle]
You are given a structured experimental record from {domain}. The record contains fixed conditions and one explore field.

Your task is to decide whether this record should be kept as a final setup-level retrieval query candidate. If it should be kept, rewrite it as one concise retrieval question.

The fixed conditions define the setup that must remain fixed. The explore field is the experimental detail being asked about. The question should target matching experimental evidence at the setup level, such as a method paragraph, setup description, or protocol detail. Do not phrase it as asking for papers, articles, studies, documents, or literature.

Follow these steps:

1. Check whether the fixed conditions define a coherent and self-contained experimental setup. The fixed fields must work together as one meaningful experimental target. Do not judge fields one by one in isolation.

2. Check whether the explore field is appropriate under those fixed conditions. The explore field should be a meaningful missing detail to ask for. It should not be already fixed, strongly implied, too weakly supported by the fixed setup, or disconnected from the fixed conditions.

3. Skip the record if it does not form a suitable retrieval question.
   Common reasons include:
   - a required parent or anchor field is missing;
   - a field value appears unstable, non-standard, or does not fit the field meaning;
   - the fixed conditions are too weak or incomplete to support the explore field;
   - the explore field is already strongly implied by the fixed conditions;
   - the field combination is methodologically awkward or internally inconsistent;
   - a numeric value, range, ratio, or loading cannot be interpreted without missing context;
   - the resulting question would be too broad, trivial, redundant, or dominated by minor protocol details;
   - the record would require adding unstated assumptions to make the question meaningful.

4. If the record is usable, write one retrieval question. Preserve all fixed conditions exactly. Ask directly for the explore field. Keep the explore field open; do not answer it or treat it as already known. You may reorder information and use exact equivalents, such as standard abbreviations, standard names, formula variants, exact unit conversions, and light syntactic changes. Do not drop fixed constraints. Do not add missing conditions. Do not broaden or narrow the setup. Do not change the meaning of numbers, ranges, or negations. If a fixed condition excludes one value of the explored field, ask naturally what is used instead under the stated conditions. If an exact reformulation is uncertain, keep the original wording.

Return only one JSON object.

If the record should be kept:
{{"request_query": "<one concise setup-level retrieval question>", "skip": false, "skip_note": null}}

If the record should be skipped:
{{"request_query": null, "skip": true, "skip_note": "<one brief sentence explaining why the record is not suitable>"}}

Do not output any explanation, note, or markdown fence.

Field definitions: {field_definitions_pretty}

Examples: {Examples}

Structured record: {record_pretty_json}
\end{lstlisting}
    \end{minipage}
    \\
    \bottomrule
    \end{tabular}
    \end{adjustbox}
    
\end{table*}
\begin{table*}[p]
    \centering
    \small
    \caption{
    Prompt for query quality filtering.
    }
    \label{prompt:query_quality_filtering}
    \begin{adjustbox}{max width=1\linewidth}
    \begin{tabular}{p{\linewidth}}
    \toprule
    \begin{minipage}{0.98\linewidth}
\begin{lstlisting}[style=promptstyle]
You are evaluating a generated query against the structured experimental record it came from. Your task is to judge whether the generated query preserves the intended meaning, is clear enough to read, and is strong enough to serve as a final benchmark retrieval query for {domain}.

Treat the structured record as the source of truth. The generated query must preserve the same experimental intent. If the record contains an explore field, the generated query must keep all fixed conditions intact and ask directly for that explore field. If the record does not contain an explore field, the generated query must function as a setup-level request for finding matching experimental evidence under the stated conditions.

Evaluate the query in the following order.

Step 1: Judge faithfulness. A faithful query must preserve the same target, all fixed constraints, all negations, all numeric values, all ranges, all units, and the same explore field if one is present. Set faithfulness = 0 if the query drops a condition, adds an unstated condition, weakens a negation, broadens a concrete value into a looser category, narrows the target, shifts a numeric boundary, changes a unit meaning, replaces the explore field with a different target, or otherwise changes the retrieval scope. Otherwise set faithfulness = 1.

Step 2: Judge clarity. A clear query should read like one interpretable setup-level retrieval request. It does not need to be elegant, but it should not be confusing, mechanically stitched together, overly awkward, title-like, or hard to understand. For records without an explore field, the query should be a complete natural-language retrieval request, preferably starting with "Find". Do not penalize minor wording differences if the request is still clear and natural. For records with an explore field, the query may be written as a direct question or as a "Find" request, as long as it clearly asks for the explore field under the fixed conditions. Set clarity = 1 if the query is understandable and reasonably well formed. Set clarity = 0 if the wording is confusing, awkward enough to reduce readability, title-like rather than request-like, or unclear about what evidence should be retrieved.

Step 3: Judge retrieval usefulness. A useful query should be specific, meaningful, and motivated enough to retrieve concrete setup-level experimental evidence. Set retrieval_usefulness = 0 if the query is too broad, too weak, too generic, too trivial, too unnatural as a real search request, or if the fixed context is not strong enough to support the target being asked for. Also set retrieval_usefulness = 0 if the open target is already strongly implied by the fixed conditions, if an important numeric or technical condition is not meaningfully anchored, if a field value appears semantically unstable or does not fit the field meaning, or if the field combination remains scientifically awkward even though the wording is readable. Otherwise set retrieval_usefulness = 1.

Step 4: Write a short diagnostic reason. The reason should identify the main issue without restating the whole record. If all three scores are 1, briefly state that the query preserves the structured intent and is clear and useful. Do not rewrite, repair, or improve the query. The evaluator should only assign scores and write a short diagnostic reason. A query should be kept downstream only when faithfulness = 1, clarity = 1, and retrieval_usefulness = 1. If any score is 0, the query should not be kept.

Return exactly one JSON object in the following format and nothing else:

{{
  "faithfulness": 0 or 1,
  "clarity": 0 or 1,
  "retrieval_usefulness": 0 or 1,
  "reason": "<short diagnostic explanation>"
}}

Examples: {Examples}
Field definitions: {field_definitions_pretty}
Structured record: {record_pretty_json}
Generated query: {generated_query}
\end{lstlisting}
    \end{minipage}
    \\
    \bottomrule
    \end{tabular}
    \end{adjustbox}
    
\end{table*}

\begin{table*}[p]
    \centering
    \small
    \caption{
    Prompt for fixed-constraint relevance judging.
    }
    \label{prompt:fixed_constraint_relevance_judging}
    \begin{adjustbox}{max width=1\linewidth}
    \begin{tabular}{p{\linewidth}}
    \toprule
    \begin{minipage}{0.98\linewidth}
\begin{lstlisting}[style=promptstyle]
You are given a structured requirement and a candidate method paragraph.

Decide whether the paragraph contains at least one coherent experimental setup that satisfies all requirements.

A coherent setup is one concrete experimental configuration. If the paragraph describes multiple variants, groups, materials, conditions, treatments, or measurement configurations, judge them separately. Do not combine evidence from different setups.

A setup satisfies the requirement only if all constraints are supported within the same setup. The support may be directly stated or directly recovered using short, evidence-grounded scientific reasoning. Do not guess missing information.

For numeric constraints, the operator "=" means numeric compatibility, not exact textual equality. Compatible unit conversions are allowed. A numeric point value and a numeric range match whenever they overlap: a required point is satisfied by a paragraph range if the point lies inside the range, and a required range is satisfied by a paragraph point if the point lies inside the required range. A continuous range reported in the paragraph is an explicit numeric value range for matching purposes. For example, 8% matches 8-15%, 40 ug matches about 30-60 ug, 37C matches 35-39C, pH 7.4 matches pH 7.2-7.6, and 0.1 M matches 100 mM. Do not reject a numeric constraint only because the paragraph reports an overlapping range rather than the exact point value.

Do not confuse continuous ranges with lists of alternative point values. A range such as 4-20% or 20-60 ug supports values inside the interval. A list such as 6%, 10%, or 12.5% only supports the listed values. If different values are assigned to different variants, targets, materials, groups, or conditions, keep the assignment within the same setup and do not infer an unstated pairing.

Set paragraph_verdict to "satisfy" if at least one setup satisfies all constraints. Otherwise set it to "violate".

Return JSON only:

{{
  "paragraph_verdict": "satisfy or violate",
  "supporting_setup": "brief description of the satisfying setup, or null",
  "evidence": [
    "short evidence span"
  ],
  "reason": "brief reason"
}}

Structured requirement:
{constraints_json}

Relevant field definitions:
{field_definitions_json}

Candidate paragraph:
{paragraph_text}
\end{lstlisting}
    \end{minipage}
    \\
    \bottomrule
    \end{tabular}
    \end{adjustbox}
    
\end{table*}
\begin{table*}[p]
    \centering
    \small
    \caption{
    Prompt for exploration relevance judging.
    }
    \label{prompt:exploration_relevance_judging}
    \begin{adjustbox}{max width=1\linewidth}
    \begin{tabular}{p{\linewidth}}
    \toprule
    \begin{minipage}{0.98\linewidth}
\begin{lstlisting}[style=promptstyle]
You are given a structured requirement, an explored field, and a candidate method paragraph.

Decide whether the paragraph contains at least one coherent experimental setup that satisfies the requirement and provides a determinate value for the explored field.

A coherent setup is one concrete experimental configuration. If the paragraph describes multiple variants, groups, materials, conditions, treatments, or measurement configurations, judge them separately. Do not combine evidence from different setups.

A setup satisfies the requirement only if all constraints are supported within the same setup and the explored field has a determinate value in that same setup. The support may be directly stated or directly recovered using short, evidence-grounded scientific reasoning. Do not guess missing information.

For categorical "=" constraints, the required value must be supported. For categorical "!=" constraints, the setup must support a concrete value different from the forbidden value. Missing or unclear values do not satisfy a "!=" constraint.

For numeric constraints, the operator "=" means numeric compatibility, not exact textual equality. Compatible unit conversions are allowed. A numeric point value and a numeric range match whenever they overlap: a required point is satisfied by a paragraph range if the point lies inside the range, and a required range is satisfied by a paragraph point if the point lies inside the required range. A continuous range reported in the paragraph is an explicit numeric value range for matching purposes. For example, 8% matches 8-15%, 40 ug matches about 30-60 ug, 37C matches 35-39C, pH 7.4 matches pH 7.2-7.6, and 0.1 M matches 100 mM. Do not reject a numeric constraint only because the paragraph reports an overlapping range rather than the exact point value.

Do not confuse continuous ranges with lists of alternative point values. A range such as 4-20% or 20-60 ug supports values inside the interval. A list such as 6%, 10%, or 12.5% only supports the listed values. If different values are assigned to different variants, targets, materials, groups, or conditions, keep the assignment within the same setup and do not infer an unstated pairing.

Set paragraph_verdict to "satisfy" if at least one coherent setup satisfies all constraints and provides the explored field value. Otherwise set it to "violate".

Return JSON only. The value of paragraph_verdict must be either "satisfy" or "violate".

{{
  "paragraph_verdict": "satisfy",
  "supporting_setup": "brief description of the satisfying setup, or null if paragraph_verdict is violate",
  "explored_field_value": "value of the explored field in the satisfying setup, or null if paragraph_verdict is violate",
  "evidence": [
    "short evidence span"
  ],
  "reason": "brief reason"
}}

For paragraph_verdict = "satisfy", provide a brief supporting_setup, the explored_field_value, and short evidence spans from the same setup. For paragraph_verdict = "violate", set supporting_setup and explored_field_value to null; evidence may be empty.

Explored field: {explore_field}

Structured requirement: {constraints_json}

Relevant field definitions: {field_definitions_json}

Candidate paragraph: {paragraph_text}
\end{lstlisting}
    \end{minipage}
    \\
    \bottomrule
    \end{tabular}
    \end{adjustbox}
    
\end{table*}

\begin{table*}[t]
\centering
\footnotesize
\setlength{\tabcolsep}{4pt}
\renewcommand{\arraystretch}{1.12}
\caption{
Prompts used for reranking experiments.
The placeholder \{query\} denotes the original ProSearch query, \{paragraph\} denotes a candidate method paragraph, and \{passages\} denotes a list or group of candidate paragraphs.
}
\label{tab:reranker_prompts}
\begin{tabularx}{\textwidth}{@{}p{0.2\textwidth}X@{}}
\toprule
Model & Prompt \\
\midrule

BGE-Reranker-v2-m3
&
\begin{minipage}{0.98\linewidth}
No natural-language prompt is added.
We use the official cross-encoder scoring interface with paired inputs:

Query: \{query\}

Passage: \{paragraph\}

The model returns a relevance score for each query--paragraph pair.
\end{minipage}
\\

\midrule
Rank-R1
&
\begin{minipage}{0.98\linewidth}
System:

A conversation between User and Assistant.
The user asks a question, and the Assistant solves it.
The assistant first thinks about the reasoning process and then provides the answer.

User:

Given the query: ``\{query\}'', which of the following documents is most relevant?

\{passages\}

After completing the reasoning process, provide only the label of the most relevant document to the query, enclosed in square brackets.
For example, if the third document is the most relevant, the answer should be: [3].

The official setwise implementation compares a set of candidates and uses the selected label during sorting.
\end{minipage}
\\

\midrule
Rank1
&
\begin{minipage}{0.98\linewidth}
Determine if the following passage is relevant to the query.
Answer only with ``true'' or ``false''.

Query: \{query\}

Passage: \{paragraph\}

The model first generates a reasoning trace and then makes a binary relevance judgment.
We use the probability of the ``true'' token as the reranking score.
\end{minipage}
\\

\midrule
DIVER-GroupRank-7B
&
\begin{minipage}{0.98\linewidth}
System:

Your task is to evaluate and rank documents based on how well they help answer the given query.
Usefulness and helpfulness are the primary criteria, followed by topical relevance.
For each document, assess how directly it addresses the user's intent and how much actionable information it provides.

User:

I will provide you \{TOPK\} documents, each indicated by a numerical identifier.

Score these documents based on their usefulness and relevance to the query.

Query: \{query\}

Documents:

\{passages\}

The output should contain a brief reasoning process followed by JSON scores, with document identifiers as keys and integer scores from 0 to 10 as values.
\end{minipage}
\\

\midrule
ReasonRank-8B
&
\begin{minipage}{0.98\linewidth}
System:

You are RankLLM, an intelligent assistant that can rank passages based on their relevance to the query.
Given a query and a passage list, first think about the reasoning process and then provide the reranked passage list.

User:

I will provide you with \{N\} passages, each indicated by a numerical identifier.
Rank the passages based on their relevance to the search query: \{query\}.

\{passages\}

Search Query: \{query\}.

Rank the \{N\} passages above based on their relevance to the search query.
All passages should be included and listed using identifiers in descending order of relevance.
The format of the answer should be [] $>$ [], e.g., [2] $>$ [1].
\end{minipage}
\\

\midrule
Qwen3-Reranker
&
\begin{minipage}{0.98\linewidth}
System:

Judge whether the document meets the requirements based on the query and the instruction provided.
The answer can only be ``yes'' or ``no''.

User input:

Instruct: Given a web search query, retrieve relevant passages that answer the query

Query: \{query\}

Document: \{paragraph\}

We use the probability of the ``yes'' token as the reranking score.
\end{minipage}
\\

\bottomrule
\end{tabularx}
\end{table*}

\begin{table*}[p]
    \centering
    \small
    \caption{
    Expert annotation instructions for query quality evaluation.
    Experts evaluate whether each generated query faithfully preserves the structured constraints, is clearly expressed, and forms a meaningful protocol retrieval request.
    }
    \label{tab:expert_query_annotation_instruction}
    \begin{adjustbox}{max width=1\linewidth}
    \begin{tabular}{p{\linewidth}}
    \toprule
    \begin{minipage}{0.98\linewidth}
\begin{lstlisting}[style=promptstyle]
Dear expert,

Thank you for helping us evaluate the quality of protocol retrieval queries in ProSearch. ProSearch is a benchmark for retrieving method paragraphs that match concrete experimental protocol requirements. Each query is generated from a structured set of protocol constraints, such as the experimental target, input material, method, and condition parameters. Your feedback will help us verify whether the generated queries are faithful to the structured constraints, clearly expressed, and useful as realistic protocol-level retrieval requests.

Each annotation case comes from your own area of expertise. For each case, we provide the relevant field definitions, the structured protocol constraints, an explored field if the query asks for a missing protocol detail, and the generated natural-language query. Please do not answer the query or rewrite it. The goal is to judge whether the query is suitable for inclusion in the benchmark.

Please evaluate each query using the following three criteria.

1. Faithfulness

Faithfulness measures whether the generated query preserves the intended meaning of the structured constraints. A faithful query should keep all fixed requirements, including entities, materials, methods, conditions, numeric values, ranges, units, and negations. If an explored field is provided, the query should ask for that field while keeping all fixed requirements unchanged. Please mark Faithfulness = Yes if the query preserves the structured intent. Please mark Faithfulness = No if it drops a requirement, adds an unstated requirement, changes a value, weakens a negation, changes a numeric value or unit, or asks for a different field.

2. Clarity

Clarity measures whether the query can be understood as a protocol retrieval request. The query does not need to be stylistically perfect, but it should be natural enough for a researcher to understand what kind of method evidence should be retrieved. For a fixed-constraint query, the query should ask for experiments, protocols, conditions, or method evidence satisfying the stated requirements. For an exploration query, it should clearly ask for the explored field under the fixed conditions. Please mark Clarity = Yes if the query is understandable and has no obvious ambiguity. Please mark Clarity = No if it is confusing, mechanically stitched together, title-like rather than request-like, or unclear about the intended evidence.

3. Retrieval Usefulness

Retrieval usefulness measures whether the query describes a meaningful protocol retrieval need. A useful query should be specific enough to retrieve concrete method-level evidence, and the fixed conditions should provide sufficient experimental context. The query should not be too broad, too trivial, scientifically incoherent, or dominated by a minor protocol detail. For exploration queries, the explored field should be a meaningful missing protocol detail rather than something already implied by the fixed conditions or disconnected from them. Please mark Retrieval Usefulness = Yes if the query is meaningful for protocol evidence retrieval. Please mark Retrieval Usefulness = No if it is too generic, underspecified, scientifically awkward, or not useful as a realistic retrieval request.

For each query, please provide Faithfulness, Clarity, and Retrieval Usefulness as Yes or No. Please also provide a short comment, especially when any answer is No. A query is considered acceptable only when all three criteria are marked Yes.

Thank you for your contribution.
\end{lstlisting}
    \end{minipage}
    \\
    \bottomrule
    \end{tabular}
    \end{adjustbox}
    
\end{table*}

\begin{table*}[p]
    \centering
    \small
    \caption{
    Expert annotation instructions for relevance label verification.
    Experts verify whether each assigned paragraph-level relevance label is correct under the protocol-level relevance definition.
    }
    \label{tab:expert_label_annotation_instruction}
    \begin{adjustbox}{max width=1\linewidth}
    \begin{tabular}{p{\linewidth}}
    \toprule
    \begin{minipage}{0.98\linewidth}
\begin{lstlisting}[style=promptstyle]
Dear expert,

Thank you for helping us verify relevance labels in ProSearch. ProSearch evaluates whether retrieval systems can find method paragraphs that satisfy concrete experimental protocol requirements. In this annotation task, your goal is to check whether the assigned relevance label for a query-paragraph pair is correct.

Each annotation case comes from your own area of expertise. For each case, we provide the structured protocol constraints, an explored field if the query asks for a missing protocol detail, the relevant field definitions, a candidate method paragraph, and the assigned relevance label. Please do not judge relevance by broad topical similarity alone. A paragraph is relevant only if it contains at least one coherent experimental setup that satisfies the protocol requirement.

Please verify each relevance label using the following steps.

1. Identify coherent experimental setups

First, read the candidate paragraph and identify the experimental setup or setups described in it. A setup means one concrete experimental configuration, such as a specific culture condition, assay setting, electrochemical measurement, synthesis route, material processing condition, or polymer formulation. If the paragraph describes multiple variants, materials, treatment groups, reaction conditions, or measurement settings, please judge them separately. Do not combine evidence from different setups.

2. Check the fixed constraints

For each coherent setup, check whether all structured constraints are supported within that same setup. A constraint may be supported by a direct statement or by short, evidence-grounded scientific reasoning. 

For categorical equality constraints, the required value must be supported in the setup. For categorical negation constraints, the setup must support a concrete value different from the forbidden value, and a missing or unclear value does not satisfy a negation constraint. For numeric constraints, compatible unit conversions are allowed. Numeric overlap is also allowed: a required point value is satisfied by a reported range if the point lies inside the range, and a required range is satisfied by a reported point if the point lies inside the required range. Please do not confuse a continuous range with a list of alternative point values. 

3. Check the explored field when applicable

If an explored field is provided, the paragraph is relevant only when the same setup satisfies all fixed constraints and also provides a determinate value for the explored field. The explored value must be recoverable from the paragraph. If the explored field is missing, ambiguous, or only mentioned in another setup, the paragraph should be marked as non-relevant.

4. Decide whether the assigned label is correct

Please mark the correct label as Relevant if at least one coherent setup satisfies all fixed constraints. For exploration queries, the same setup must also provide the explored field value. Please mark the correct label as Non-Relevant otherwise. Then compare your correct label with the assigned label. If they match, mark Label Correct = Yes. If they do not match, mark Label Correct = No and provide the corrected label.

For each query-paragraph pair, please provide the correct label, whether the assigned label is correct, a brief description of the supporting setup if the paragraph is relevant, short evidence span or spans supporting your decision, and a short comment explaining your judgment.

Thank you for your contribution.
\end{lstlisting}
    \end{minipage}
    \\
    \bottomrule
    \end{tabular}
    \end{adjustbox}
    
\end{table*}


\clearpage
\bibliographystyle{plain}
\begin{small}
    \bibliography{reference}
\end{small}

\end{document}